\documentclass{aa} 
\usepackage{graphicx}
\usepackage{txfonts}
\usepackage[allcolors=blue]{hyperref}

\graphicspath{{./}{FIG_FINAL/}}

\usepackage{adjustbox}

\usepackage[normalem]{ulem}

\begin{document}

\title{Photometric binaries, mass functions, and structural parameters of 78 Galactic open clusters\footnote{Tables~\ref{tab:tab1} and \ref{tab:tab2} are only available in electronic form
at the CDS via anonymous ftp to \url{cdsarc.cds.unistra.fr}}}

\author{
    Giacomo Cordoni \inst{1}$^{\and}$ \inst{2} \and Antonino P. Milone \inst{1}$^{\and}$ \inst{2} \and Anna F. Marino \inst{2}$^{\and}$ \inst{3} \and Enrico Vesperini \inst{4} \and Emanuele Dondoglio \inst{1} \and Maria Vittoria Legnardi \inst{1} \and Anjana Mohandasan \inst{1} \and Marilia Carlos \inst{1} \and Edoardo P. Lagioia \inst{1}  \and Sohee Jang \inst{5} \and Tuila Ziliotto \inst{1}
}
\institute{
    Dipartimento di Fisica e Astronomia ``Galileo Galilei'', Univ. di Padova, Vicolo dell'Osservatorio 3, I-35122 Padova, Italy
    \and
    Istituto Nazionale di Astrofisica - Osservatorio Astronomico di Padova, Vicolo dell’Osservatorio 5, IT-35122, Padova, Italy
    \and 
    Istituto Nazionale di Astrofisica - Osservatorio Astrofisico di Arcetri, Largo Enrico Fermi, 5, Firenze, IT-50125, Firenze, Italy
    \and 
    Department of Astronomy, Indiana University, Bloomington, IN 47401, USA
    \and
    Center for Galaxy Evolution Research and Department of Astronomy,Yonsei University, Seoul 03722, Korea
}
\abstract
{Binary stars play a crucial role in our understanding of the formation and evolution of star clusters and their stellar populations}
{We use Gaia Data Release 3 to homogeneously analyze 78 Galactic open clusters and the unresolved binary systems they host, each composed of two main sequence (MS) stars.}
{We first investigated the structural parameters of these  clusters, such as the core radius and the central density, and determined the cluster mass function (MF) and total mass by  interpolating
the density profile of each cluster. We measured the fraction of binaries with a large mass ratio and the fraction of blue straggler stars (BSSs), and finally investigated possible connections between the populations of binary stars and BSSs with the main parameters of the host cluster.}
{Remarkably, we find that the MFs of 78 analyzed open clusters follow a similar trend and are well reproduced by two single power-law functions, with a change in slope around masses of 1$M_{\odot}$. The fraction of binary stars ranges from $\sim$15\% to more than $\sim$60\% without significant correlation with the mass and the age of the host cluster. Moreover, we detect hints of a correlation between the total fraction of binary stars and the central density of the host cluster. We compared the fraction of binary stars with that of BSSs, finding that clusters with high and low central density exhibit different trends.
The fraction of binaries does not significantly change with the mass of the primary star and the mass ratio. The radial distribution of binary stars depends on cluster age. The binaries of clusters younger than $\sim$800\,Myr typically show a flat radial distribution, with some hints of a double peak. In contrast, the binaries of the remaining clusters are more centrally concentrated than the single stars, which is similar to what is observed in globular clusters.}
{}

\keywords{Star clusters, Open clusters,  binary stars, mass functions, structural parameters}
\titlerunning{Binaries in Open Clusters}
\authorrunning{Cordoni et al.}

\maketitle

\section{Introduction} \label{sec:intro}

Characterization of binary stellar systems in star clusters is a crucial step in shedding light on various fields of stellar astrophysics, including the dynamic evolution of stellar systems,    star formation, and stellar evolution. 
For example, a robust determination of the physical properties of a star cluster, including the mass function (MF) and the total mass, would require significant knowledge of its populations of binary stars.
Furthermore, the stellar evolution of a binary system can strongly differ  from that of single stars, and depends on the physical properties of the system, including the mass ratio, binding energy, and orbital period. 
\\

Various approaches can be used to identify and characterize  binary stars in stellar clusters. For example, binaries can be detected from radial-velocity variation or from photometric variability. While these methods have the advantage of constraining each binary system, they are either limited to bright stars or biased toward binaries with short periods.  \\

In this work, we follow an alternative approach based on the fact that binary stars formed by couples of main sequence (MS) stars exhibit redder colors with respect to single MS stars. Hence, binary stars can be identified as stars lying on the red side of the MS fiducial line in the color--magnitude diagram \citep[CMD; e.g.,][]{romani1991a, bolte1992a, rubenstein1997a, bellazzini2002a, clark2004a, richer2004a, zhao2005a, milone2009a}. The main advantages of this approach are that \textit{(i)} it requires observations in only two different filters, hence requiring a small amount of telescope time; (\textit{ii)} it allows us to simultaneously investigate large numbers of stars in the CMD; and (\textit{iii)} the detection efficiency does not depend on binary properties such as period and inclination. \\
Clearly, high-precision photometry is needed to disentangle binaries from single stars. Moreover, the correction for differential reddening and the identification of field stars that contaminate the cluster CMD are crucial ingredients to infer the fraction of binaries from photometry. On the other hand, the use of this method comes with some caveats, as follows.

This approach has been used to investigate the binaries of a large sample of 67 Galactic GCs \citep[see e.g.,][]{sollima2007a, milone2012a, ji2015a, milone2016a} using homogeneous photometry from images collected with the {\it Hubble Space Telescope}, and for Galactic open clusters with the combination of multiple surveys \citep{malofeeva2022, malofeeva2023}.
Photometric binaries have been widely investigated in in Galactic open clusters (OCs), although most studies published so far are limited to a smaller number of clusters \citep[e.g.,][and references therein]{sharma2008a, sollima2010a, cordoni2018a, ebrahimi2022a}. 
Historically, the main challenge in constraining the properties of photometric binaries in open clusters was provided by the contamination from field stars, which is critical in most Galactic open clusters.

In the past few years, the Gaia mission has provided high-precision photometry, proper motions, and parallaxes of nearly 2 billion stars in the Milky Way \citep[][Gaia DR2 and DR3]{gaiaDR2, gaiaDR3}. This exquisite dataset has allowed accurate separations to be obtained  between the bulk of the OC members and the background and foreground stars \citep[e.g.,][]{cantatgaudin2018a, bossini2019a, cantatgaudin2020, dias2022a}.

In this work, we use Gaia DR3 data to investigate a sample of 78 Galactic OCs, constrain their structure parameters, and characterize their photometric binaries. 
The paper is structured as follows: Section~\ref{sec:stars selection} describes the clusters sample choice, the cluster member selection, and differential reddening corrections. In Sections~\ref{sec:parameters}, we derive the physical parameters of  clusters, while Section\,\ref{sec:bin} is dedicated to investigation of the binaries. Finally, we present and discuss the overall results in Section~\ref{sec:results}.

\section{Data and star sample selection}
\label{sec:stars selection}

\begin{figure}
    \centering
    \includegraphics[width=0.5\textwidth]{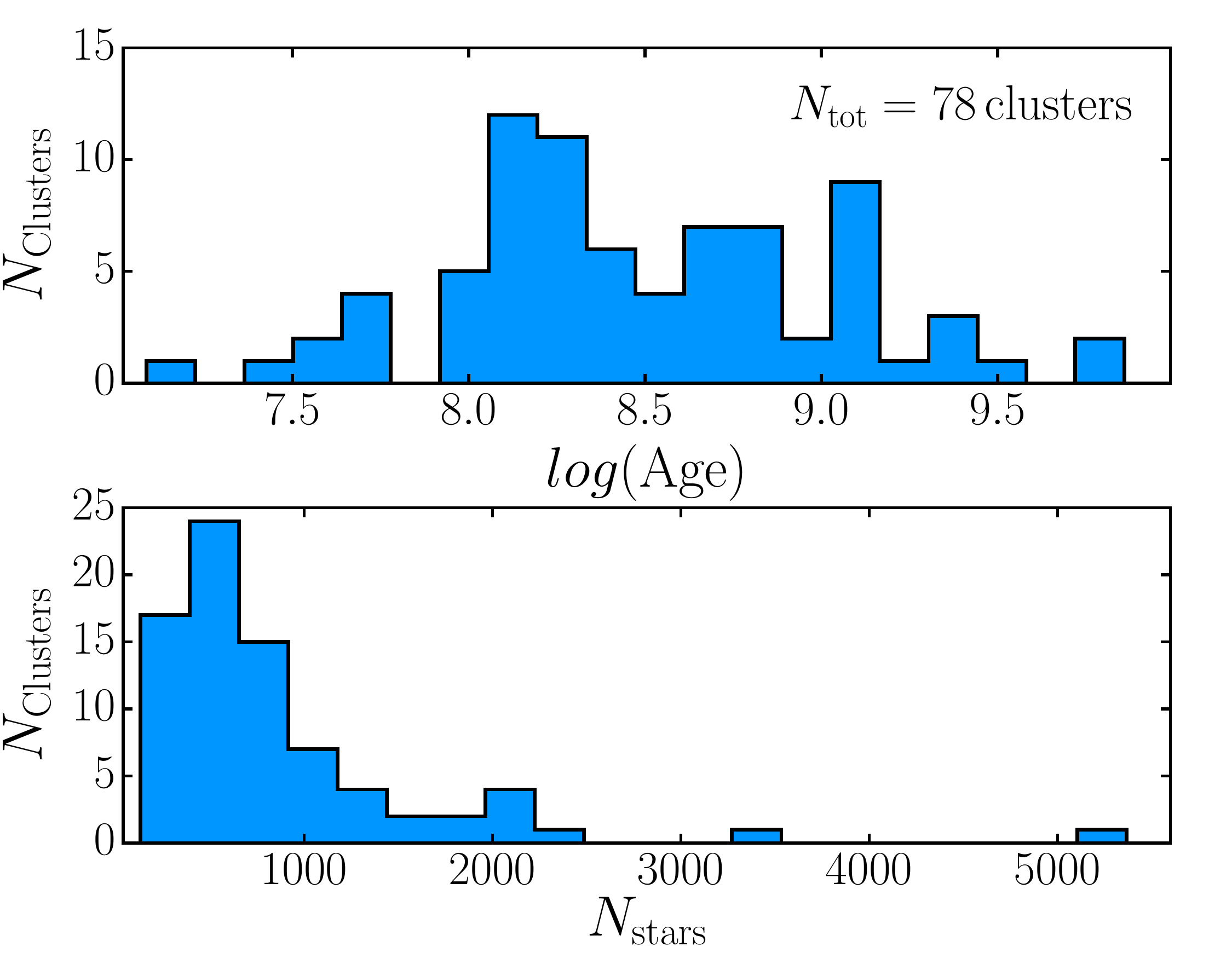}
    \caption{Properties of analyzed clusters.\textit{Top.} Histogram distribution of the ages (in logarithm scale) of the 78 open clusters  analyzed here. \textit{Bottom.} Histogram distribution of the number of cluster members. See the text for details.}
    \label{fig:par_distr}
\end{figure}

To identify unresolved binary MS-MS stars in the CMD, we need high-precision photometry, low field contamination,  negligible reddening effect, and a sufficiently large sample of stars. 
Hence, we downloaded the Gaia DR3 astro-photometric catalog of all the 269 open clusters analyzed in \citet{bossini2019a},  and selected only those clusters that satisfied our requirements. Specifically, we included only clusters with a large number of stars, that is, more than 150 cluster members, low total reddening, namely $E(B-V)\leq 0.5$, and with reasonable differential reddening variation and field-star contamination.  Specifically, as the determination of accurate binary star fractions relies on precise estimates of cluster age, distance, reddening, and metallicity, we opted to use the clusters discussed in \citet{bossini2019a} in place of larger cluster catalogs \citep[see e.g.,][]{cantatgaudin2018a, cantatgaudin2020}. Indeed, Bossini and collaborators restricted their analysis to those clusters for which the isochrone-fitting procedure would produce the most reliable age and distance estimates, thus selecting 269 out of the 1229 clusters analyzed in \citet{cantatgaudin2018a}. Including clusters with less accurate estimates could result in additional sources of uncertainty. The same clusters are also included in the more recent analysis of \citet{dias2021a}. Specifically, \citet{dias2021a} provide clusters' proper motion estimates as well as isochrones fit parameters, such as age, reddening, distance modulus and metallicity, determined through direct isochrone fitting. Hence, in the following we will make use of the cluster's properties inferred by Dias and collaborators.
Moreover, the total number of clusters analyzed in this work is 78, and their main properties are summarized in Table~\ref{tab:tab1}. 
The distributions of age and the total number of stars of the selected clusters are shown in Fig.~\ref{fig:par_distr}. Clearly, the selected clusters span a wide age range, from a few million years (e.g., $\sim\,10\,Myr$ Pozzo 1) to almost 7.5\,Gyr for NGC\,6791. Similarly, the number of cluster members varies from a minimum of 130 for NGC\,2632 to more than 5000 stars for NGC\,6705. 
Therefore, the large variety of parameters allows us to explore the behavior of binary stars in relation to different cluster structural parameters; for example, mass\footnote{The procedure adopted to infer cluster radius and cluster mass is discussed in Section~\ref{subsec:radius} \ref{subsec:mass}}, age, dimension, and other physical parameters.  \\

The identification of unresolved binary stars relies on the separation between single MS stars and MS-MS binary stars in the CMDs and requires high-precision photometry. To select a sample of well-measured cluster stars, we exploited the exquisite photometry and astrometry provided by Gaia DR3. 
We first used some Gaia DR3 quality parameters that are indicative of the quality of the photometric and astrometric measurements,  including the renormalized unit weight error (\texttt{RUWE}) and the photometric excess noise. These quantities are used to select a sample of stars with high-quality astrometry and photometry, as in \citet[][]{cordoni2018a}. 

\begin{figure*}[ht!]
    \centering
    \includegraphics[width=\textwidth]{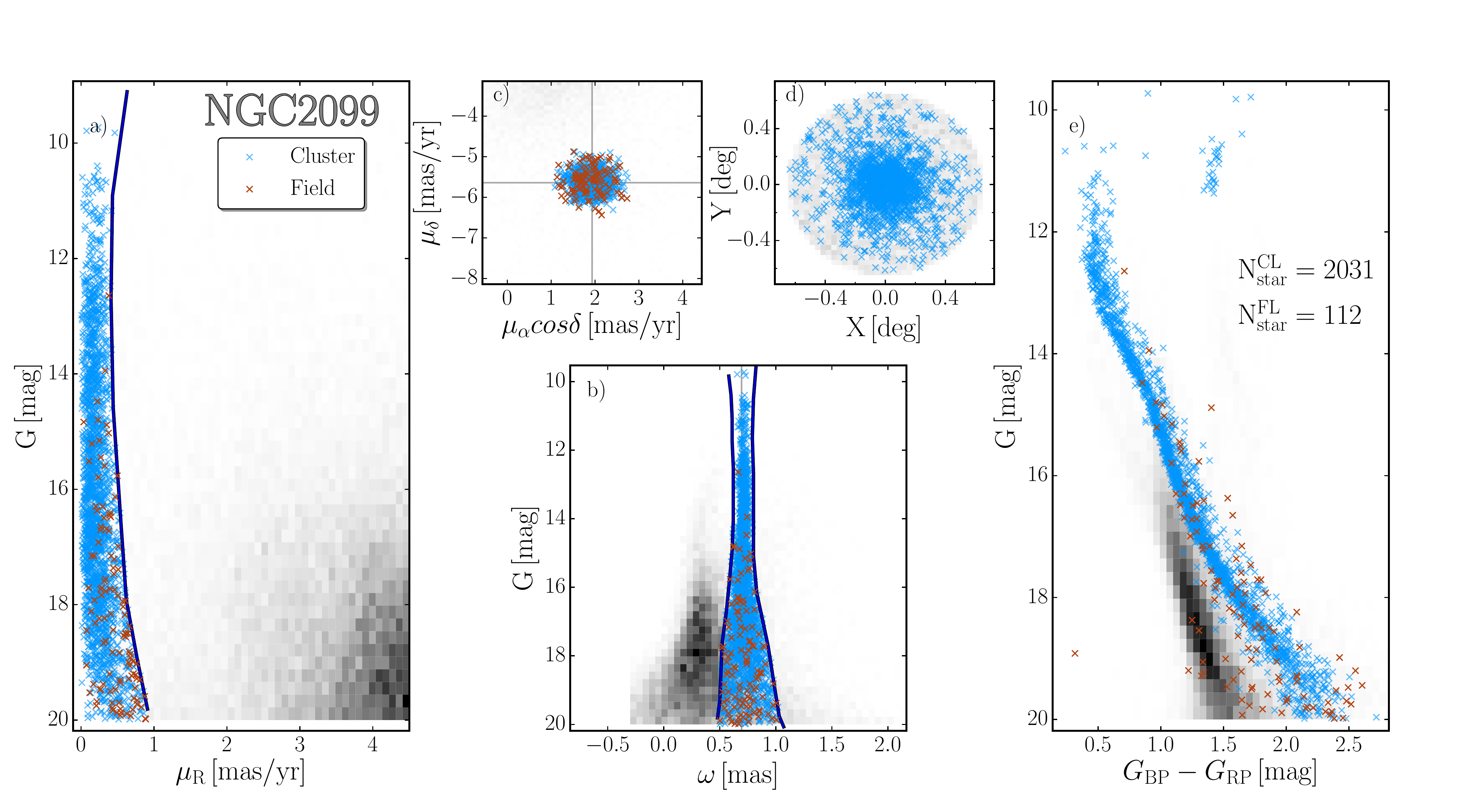}
    \caption{Cluster members selection. Procedure carried out to select the probable NGC\,2099 cluster members located within $\sim30$ arcmin from the cluster center. Panel (a): Stellar proper motions relative to the average cluster motion against the $G$ magnitude. The blue line separates stars with cluster-like proper motions from the remaining stars.  Panel (b): Stellar $G$ magnitude as a function of parallax. Stars with cluster-like parallaxes are located between the two blue lines.  The vector point diagram of proper motions is plotted in panel (c), whereas panel (d) shows the spatial stellar distribution. 
    Finally, panel (e) illustrates the $G$ vs.\,$G_{\rm BP}-G_{\rm RP}$ observed CMD, together with the best-fit isochrone. Likely cluster members are represented with azure crosses, while brown crosses indicate stars with cluster-like proper motions and parallaxes in the reference field.}
    \label{fig:selection}
\end{figure*}

To select cluster members, we adopted the approach described in \citet{cordoni2018a} and illustrated in Figure~\ref{fig:selection} for NGC\,2099. The main steps of this method can be summarized as follows.

We first analyzed the proper-motion diagram and computed the proper motion of each star relative to the cluster:
$\mu_{\rm R}=\sqrt{ (\mu_\alpha cos\delta-\left\langle \mu_\alpha cos\delta\right\rangle)^2 + (\mu_\delta-\left\langle \mu_\delta\right\rangle)^2 }$. 
Here, $\left\langle \mu_\alpha cos\delta\right\rangle$ and $\left\langle \mu_\delta\right\rangle$ are the average cluster proper motions from \citet{dias2021a}.
As an example, Figure~\ref{fig:selection}(a) shows $\mu_{\rm R}$ as a function of the G magnitude for stars in the field of view of NGC\,2099.
We divided the interval between G=9.0 and 20.0 mag into bins of 0.5 mag and calculated for each bin the median value of $\mu_{\rm R}$ and the corresponding root-square mean ($\sigma$).
To derive the blue line of Figure~\ref{fig:selection}a, we first shifted the median value of $\mu_{\rm R}$ of each bin by $3\sigma$ and associated this quantity with the average magnitude of the star in the bin. The blue line is derived by linearly interpolating these points.
We considered the stars on the left side of the blue line to be  candidate cluster members (azure crosses in Figure~\ref{fig:selection}a).
For completeness, we show the proper-motion diagram in Figure~\ref{fig:selection}c).

We then improved the selection of cluster members using stellar parallaxes (see Figure~\ref{fig:selection}b) for NGC\,2099).
Similarly to what we did in panel (a), we divided the diagram into magnitude bins and derived the median parallax of the proper-motion-selected stars and the corresponding rms ($\sigma$) of each bin.
We shifted the median parallax value of each bin by a $\pm 3\sigma$ region and associated the resulting values with the average magnitude of the stars in each of those bins. 
We derived the left and right blue lines plotted in panel (d) of Figure~\ref{fig:selection} by linearly interpolating these points.
We considered the proper-motion-selected stars between the two lines as probable cluster members and represented them with azure crosses.

Clearly, the sample of selected candidate cluster members could include residual field stars with proper motions and parallaxes similar to those of the cluster. 
To estimate the contribution of contaminating field stars, we identified a reference field located well outside the selected cluster field, that is, at a radial distance ten times larger than the radius containing 50\% of the total number of stars determined in \citet{dias2021a}.
We selected the field stars that populate the same regions of the $G$ vs.\,$\mu_{\rm R}$ and $G$ vs.\,$\omega$ diagrams populated by cluster members and that pass the selection criteria described above.
These field stars are marked with brown crosses in Figure~\ref{fig:selection}.
  
Finally, we defined the outer cluster radius as the radius where the density of cluster stars is equal to the density of residual field stars. Figure~\ref{fig:selection}(d) shows the spatial distribution of the selected NGC\,2099 stars. The CMD of the 2031 cluster members and the 112 selected  reference field stars is plotted in Figure~\ref{fig:selection}e.

\subsection{Differential reddening}
As our goal is to identify unresolved binary stars in the CMD, we need to minimize the broadening of stellar colors and magnitudes due to spatial reddening variations. To correct the photometry for differential reddening, we followed the method described in \citet{milone2012a}. 
Briefly, we started by selecting a sample of reference stars,  which is composed of bright MS cluster members, and calculated the corresponding fiducial line in the CMD.  We used the absorption coefficients by \citet{casagrande2018a} to determine the direction of the reddening and calculate the color and magnitude distance of each reference star from the fiducial along the reddening direction.
 
To calculate the differential reddening associated with each star, we selected the 25 neighboring reference stars and calculated the median distance along the reddening line. We excluded each reference star from the determination of its own differential reddening. The corresponding error is calculated as the root mean scatter of the distance values divided by the square root of $N-1$. We refer to \citet{milone2012a} for a more detailed description of the procedure. 
 
\begin{figure*}
    \centering
    \includegraphics[width=\textwidth, trim={0cm 0cm 0cm 0cm}, clip]{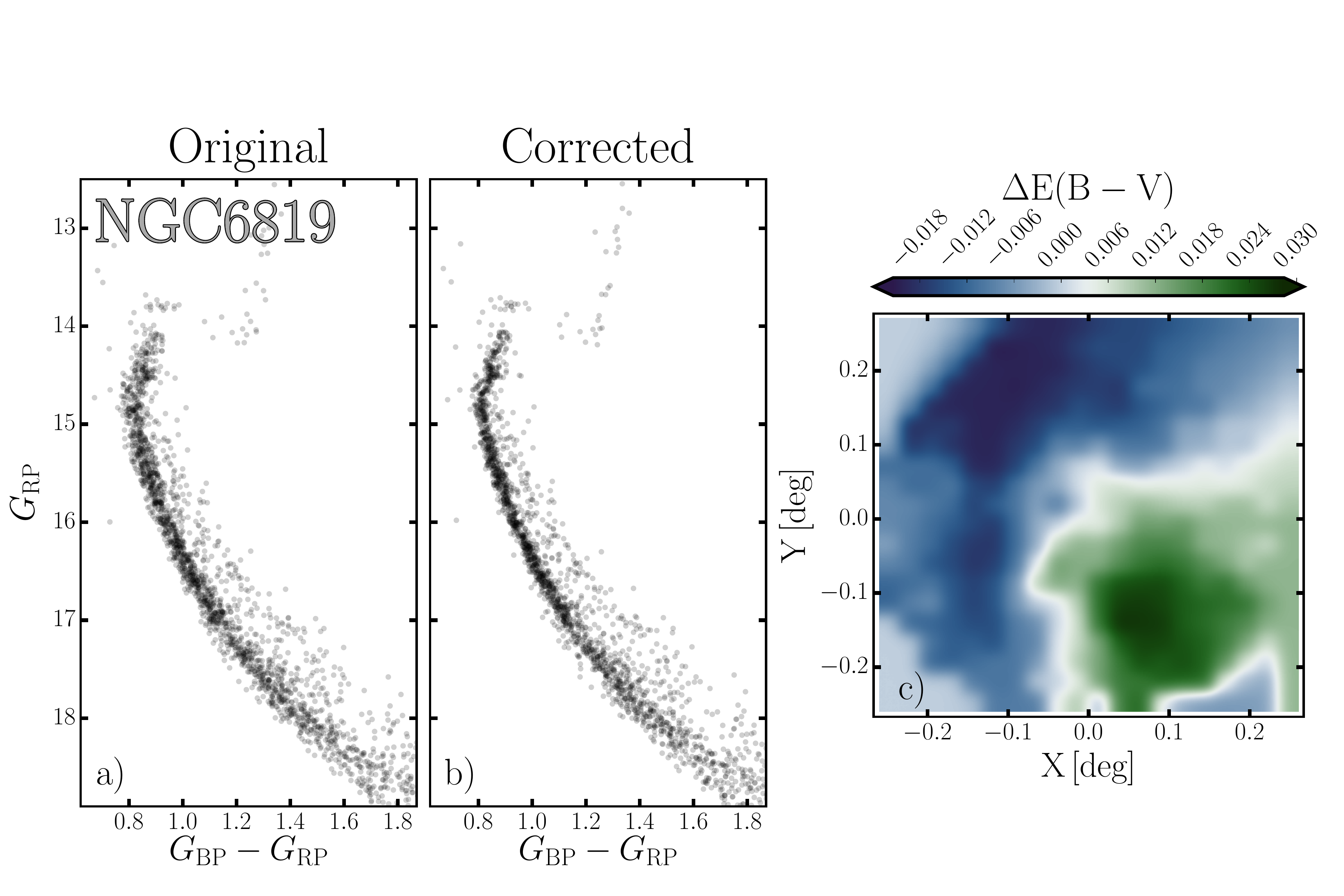}
    \caption{Differential reddening correction. Comparison of the $G_{\rm RP}$ vs.\,$G_{\rm BP}-G_{\rm RP}$ CMD before (panel a) and after the correction for differential reddening (panel b). The differential reddening map is shown in panel (c), color-coded according to the top color bar.}
    \label{fig:dr}
\end{figure*}

The results of the differential reddening correction are illustrated in Figure~\ref{fig:dr} for the open cluster NGC\,6819. Specifically, panels (a) and (b) show the original and corrected photometry, respectively, while panel (c) shows the differential reddening map where blueish colors indicate negative DR and greenish colors indicate positive variations. Results on NGC\,6819 highlight the effectiveness of the differential-reddening-correction procedure. Indeed, all the photometric sequences in the corrected CMDs are much narrower when compared to the original CMD.
     
We applied the above procedure to all the clusters of our sample. In 44 out of 78 clusters, the inferred reddening variations are significantly larger than the corresponding errors. In the following analysis, for these clusters marked with the flag \texttt{DR} in Table~\ref{tab:tab1} we make use of the photometry corrected for differential reddening. Their differential-reddening maps are available as supplementary material, together with the individual differential-reddening-correction catalogs.\\

\subsection{The color--magnitude diagrams}
\label{subsec:cmds}

The differential-reddening-corrected CMDs of the selected cluster members from the 78 clusters are shown in Figures~\ref{fig:cmds1}-\ref{fig:cmds4} of the Appendix. Clusters are sorted by age, as determined in \citet{dias2021a}, from the youngest, namely\,Pozzo\,1 (10\,Myr), to the oldest, namely NGC\,6791 (7.5\,Gyr). 
Visual inspection of these CMDs shows that at least 30-35 clusters exhibit the extended MS turn-off (eMSTO).  Such clusters are flagged with \texttt{eMSTO} in Table~\ref{tab:tab1}. 
They comprise about $\sim$45\% of the 72 studied clusters younger than 2\,Gyr, which is the limiting age for the occurrence of the eMSTO in star clusters \citep[e.g.,\,][]{milone2009a, goudfrooij2014a, niederhofer2015a, cordoni2018a}.

Our results corroborate the evidence that the eMSTO is a typical feature of young clusters. However, we note that  at least 35 clusters with ages between 0.01 and 2\,Gyr are consistent with simple populations. Hence, the eMSTO is not a ubiquitous property of these clusters. 
We refer to Cordoni et al.\,(in preparation) for further investigation of the eMSTO phenomenon.
  
\section{Physical parameters of open clusters }
\label{sec:parameters}
Accurate estimates of the structural parameters of the main clusters are required to characterize their  binary stars. In the following, we use Gaia DR3 astrometry and photometry to determine the core radius, density, and mass of each cluster.

\subsection{Density profile}
\label{subsec:radius}

To constrain the density profile of each cluster, we divided the field of view into overlapping annular bins of constant width, defined as the difference between the external and internal radius. The width has been chosen so as to maximize the spatial information while ensuring good statistics.
We calculated the stellar density in each annulus, ($\sum^*$), as the number of cluster stars brighter than $G_{\rm RP}$=18.5 mag  per square arcmin. We assumed that the Gaia DR3 catalog is complete for magnitudes brighter than $G_{\rm RP}$=18.5 as discussed in \citet[][]{boubert2020} and \citet{everall2022}. To account for residual field contamination, we determined the density of field stars, selected as described in Sect.~\ref{sec:stars selection}, and subtracted such value from the observed cluster field density. The uncertainties that we associated with stellar density are determined as Poisson errors.

We plotted $\sum^*$ as a function of the average radius of each annulus ($r$) and performed a least-square fit of the observed density profile with the function discussed in \citet[][EFF profile]{elson1987a}:

\begin{equation}
    \mu(r) = \mu_0 \left( 1+\frac{r^2}{a^2} \right)^{-\gamma/2}
    \label{eqn:eef}
,\end{equation}
where $\mu_{0}$ represents the central density in $\rm stars/arcmin^2$, $a$ a scale radius, and $\gamma$ is the index of the power law at large radial distances from the cluster center. The cluster core radius, $r_c$, is linked to $a, \gamma$ by the  relation:
\begin{equation}
    r_c = a \left( 2^{2/\gamma}-1 \right)^{1/2}
.\end{equation}

\begin{figure}
    \centering
    \includegraphics[width=0.5\textwidth, trim={0cm 0cm 0cm 0cm}, clip]{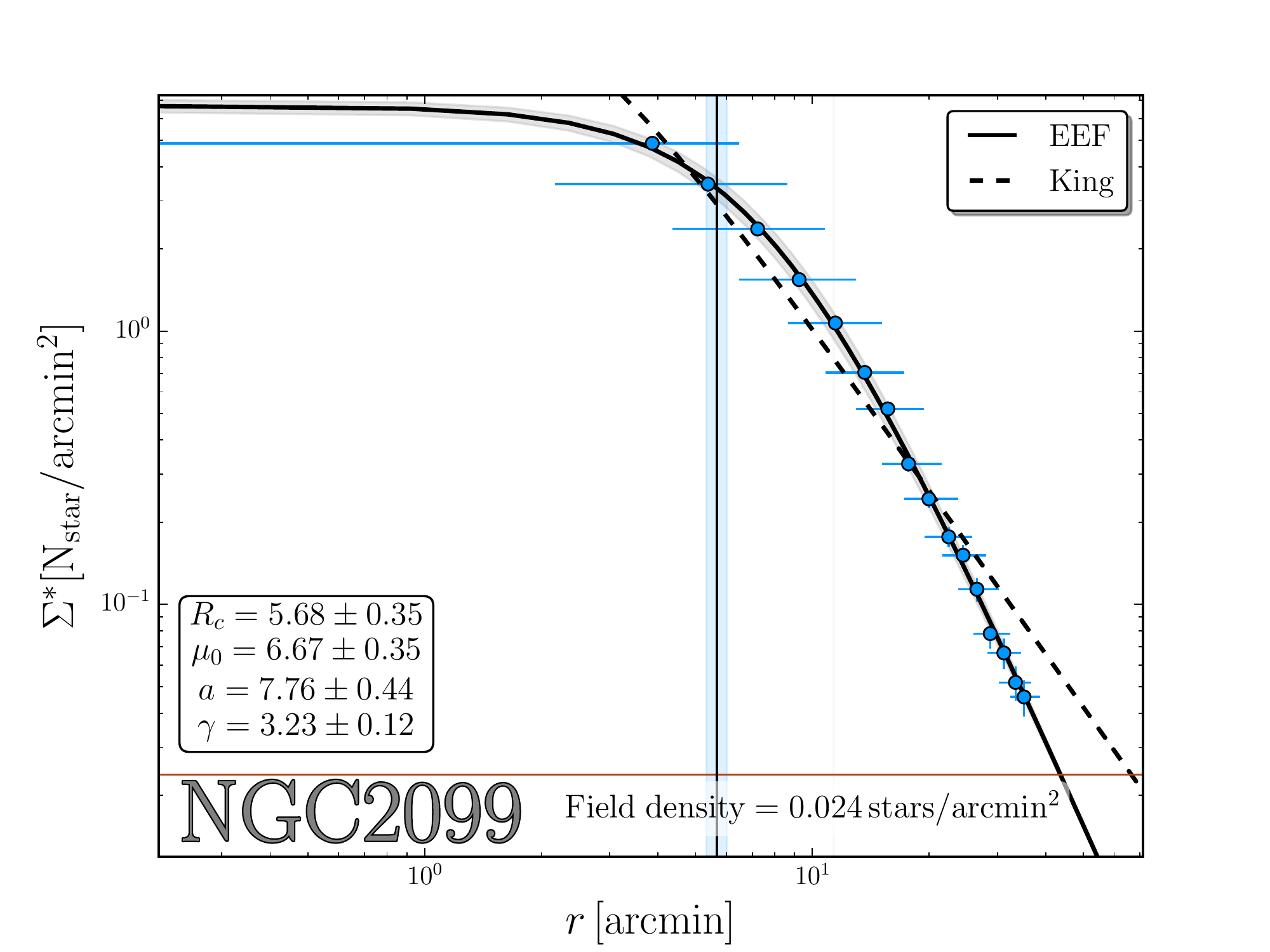}
    \caption{Density profile of NGC\,2099. Azure points with error bars represent the observed stellar density profile. 
     The black solid line indicates the best-fitting EEF \citep{elson1987a} profile, while the corresponding uncertainty is indicated by the gray shaded area. The inferred core radius is indicated with the black solid vertical line, while the shaded region is its uncertainty. The best-fitting parameters are quoted in the bottom-left inset.
     The dashed line shows the best-fitting
      King \citep{king1962a} profile.  }
    \label{fig:eef fit}
\end{figure}

The EFF profile differs from the classical King profile \citep{king1962a} for the absence of tidal truncation.
As an example, Figure~\ref{fig:eef fit} compares the best-fit EFF and King profiles (black solid and dashed lines, respectively) for the test case cluster NGC\,2099. 
We verified that the conclusions are not significantly affected by the adopted bin size. To this end, we repeated the analysis for bins of different sizes and used annuli with different radii that include the same number of cluster stars. \\
Moreover, as we limit our analysis to stars in a specific magnitude range, we lose the information about all stars that lie outside the analyzed range. Hence, the recovered central stellar density may be underestimated. To deal with this bias, we exploited the recovered MFs discussed in Section~\ref{subsec:mass} to compute the total expected number of stars in each cluster. Specifically, we multiplied the fitted central density by the ratio between the expected total number of stars inferred from the MF, and the number of stars used to compute the density profile.

\subsection{Cluster mass and mass function}
\label{subsec:mass}
To estimate the total stellar mass of each cluster, we first determined the MF of MS stars brighter than $G_{\rm RP}=18.0/18.5$ according to each specific CMD quality. In particular, we adopted $G_{\rm RP}=18.5$ in 46 clusters and $G_{\rm RP}=18.0$ in the remaining 32 clusters.
We first divided the MS into ten magnitude intervals, each containing 10\% of the total number of MS stars.
Stellar magnitudes are then converted into masses using the relations by \citet[][]{marigo2017a}. \\
As a by-product, we determined the half-mass radius $(r_{\rm h})$ interpolating the cumulative mass distribution.

\begin{figure}
    \centering
    \includegraphics[width=0.5\textwidth, trim={0cm 2cm 0cm 0cm}, clip]{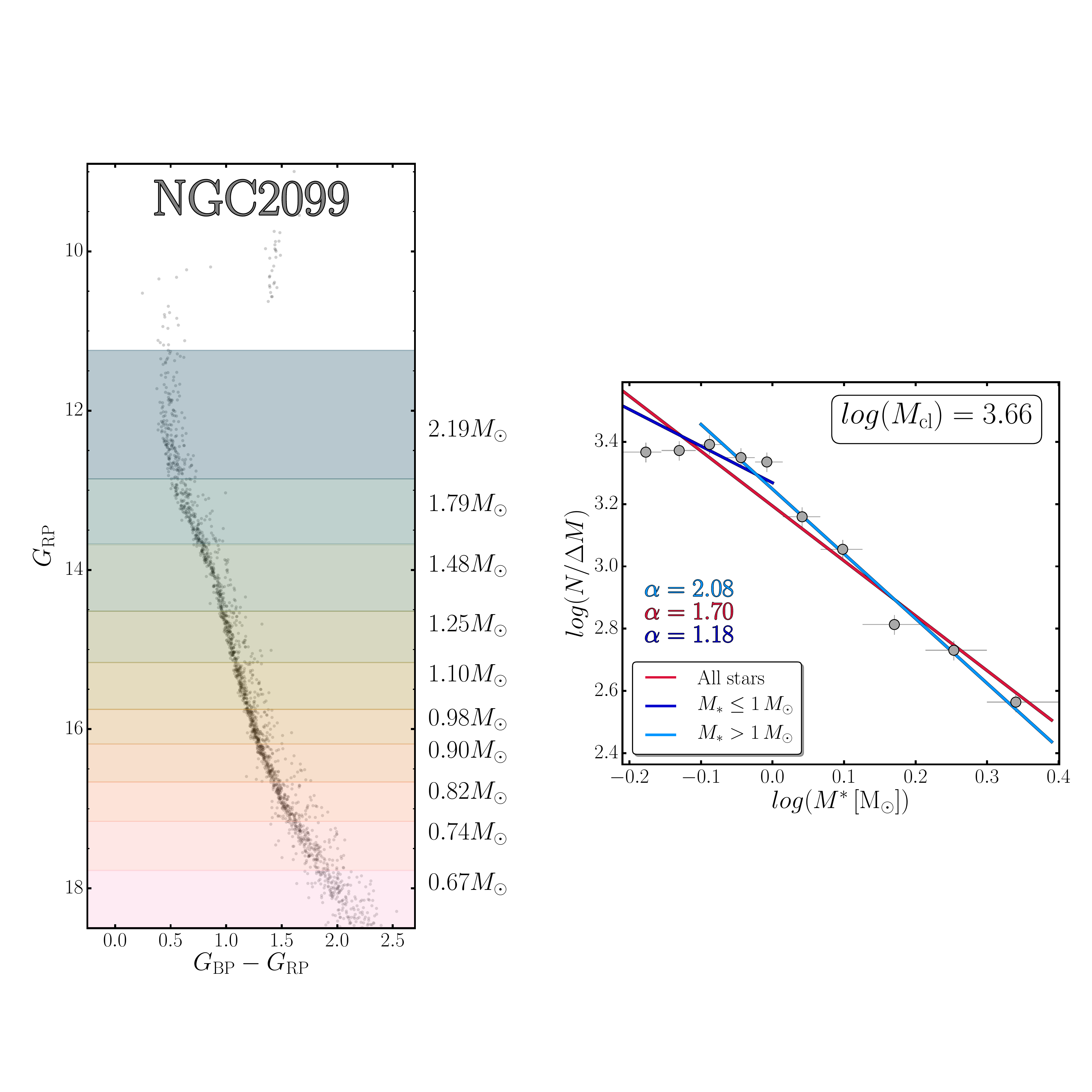}
    \caption{Mass function determination. \textit{Left panel.} $G_{\rm RP}$ vs. $G_{\rm BP}-G_{\rm RP}$ CMD of NGC\,2099. The magnitude intervals used to derive the MF are highlighted with different colors, while the median mass of each bin is indicated on the right. \textit{Right panel.} MF for the stars plotted in the left panel, i.e., $log(N/\Delta M)$ vs.\,stellar mass. The blue and azure straight lines represent the best-fit straight lines of low- ($M\leq 1M_{\odot}$) and high-mass stars ($M> 1M_{\odot}$), respectively. The red line is obtained by fitting the whole MF. The slopes are quoted in the bottom-left legend, while the total stellar mass of the cluster is indicated in the top-right inset. }
    \label{fig:mass}
\end{figure}

As an example, the different mass intervals used to infer the MF of NGC\,2099 are highlighted with different colors in the CMD of Figure~\ref{fig:mass}a, where we also indicate the mean mass of the MS stars in each bin (M).
In Figure~\ref{fig:mass}(b), we plot the logarithm of $N$ normalized to the mass interval of each bin $\Delta M$ as a function of the logarithm of M. \\
The present-day stellar MF is described by a power law with index $\alpha$:
\begin{equation}
    dN/dM = k\cdot m^{-\alpha}
    \label{eqn:mf}
,\end{equation}
with $k$ a normalization constant. Taking the logarithm of equation~\ref{eqn:mf}, we obtain the equation of a straight line with slope $\alpha$.
\begin{equation}
    log(dN/dM) = log(k)-\alpha log(m).
    \label{eqn:logmf}
\end{equation}

Hence, the observed MF can be interpolated with a straight line in $log(\Delta N/\Delta M)$ vs. $log(M)$ space by means of least squares, as in the right panel of Fig.~\ref{fig:mass} for NG\,2099. Visual inspection of the MFs of several clusters, including NGC\,2099, reveals a change in the MF slope around  $M=1\,M_{\odot}$. 
To account for this feature of the MF, we divided each cluster stellar catalog into a low-mass region with\,$M\leq 1\,M_{\odot}$, and a high-mass region with $M> 1\,M_{\odot}$, and fitted the function described by Equation\,\ref{eqn:logmf} to each of the two regions individually. Results of the three interpolations, that is, over the whole MS, just the low-mass region, and just the high-mass region, are represented with red, blue, and azure colors, respectively. We find that the change of slope is significant beyond the $3\sigma$ level in 30, or 38\%, of the 78 analyzed clusters.\\

The curves that provide the best fit to all stars, low-mass stars, and high-mass stars of NGC\,2099 are represented with red, blue, and azure lines, respectively, in Fig.~\ref{fig:mass}, where we also quote the corresponding values of $\alpha$, indicated on left. Specifically, we find an overall slope of $\alpha=1.70\pm0.17$, while the best-fit slopes of low- and high mass-stars MF are $1.18 \pm 0.04$ and $2.08 \pm 0.12$, respectively.  \\
Finally, we then used the derived overall MF slopes to infer cluster masses, which are listed in Table~\ref{tab:tab2}. Speficically, we integrated the derived MF beyond the range of observed stellar masses, including evolved stars. Implicitly, we made the reasonable assumption that the MFs do not vary beyond the stellar mass interval used to derive the MF itself. However, as evolutionary timescales after the turn-off are very short, the mass range of stars beyond the turn-off would be negligible even in the case of a varying MF. 
We derived the half-mass relaxation time $(t_{\rm rh})$  as in \citet{spitzer1987}

\begin{equation}
t_{\rm rh} = 0.138\frac{M_{tot}^{1/2}r_{\rm h}^{3/2}}{G^{1/2}\bar{m}ln(\gamma M_{\rm tot}/\bar{m})}
\label{eqn:eq 3}
,\end{equation}
where $\gamma=0.11$ and $\bar{m}$ is the mean stellar mass. The inferred cluster parameters are listed in Table~\ref{tab:tab2}.

\begin{figure*}[ht!]
    \centering
    \includegraphics[width=\textwidth, trim={0cm 0cm 0cm 0cm}, clip]{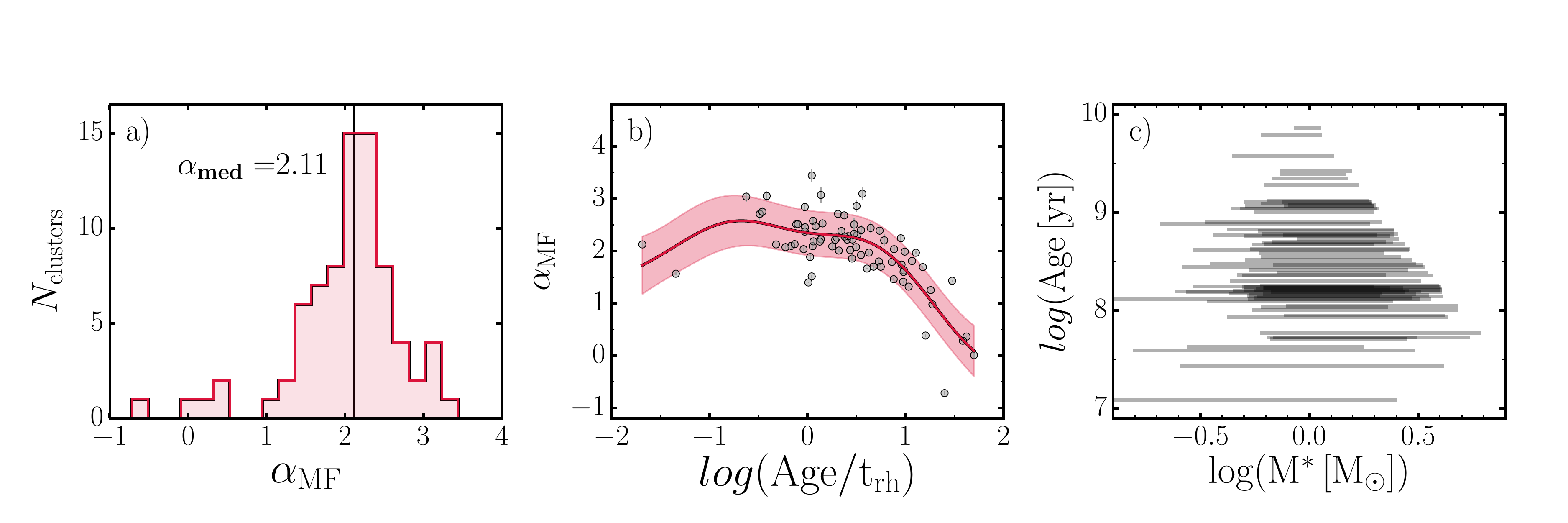}
    \caption{Mass functions properties. \textit{Panel (a):} Distribution of the overall MF slope inferred for the 78 analyzed clusters. \textit{Panel (b):} MF slope $\alpha$ vs. cluster age.  \textit{Panel (c):} Cluster age vs. explored stellar mass range in MS stars.}
    \label{fig:alpha distribution}
\end{figure*}

The results for the MF fits are summarized in Table\,\ref{tab:tab2} for the 78 analyzed clusters. As shown in Figure~\ref{fig:alpha distribution}a, we find that the MF slope distribution peaks around $\alpha \sim 2.11$ and ranges from $\sim -0.5$ to $\sim 3.5$. Moreover,  61 out of 78 clusters (78\%) are consistent with a Salpeter MF ($\alpha=2.35$) at the $1\sigma$ level.
Remarkably, we find that the MF slope depends on the dynamical age of the cluster, which is defined as the ratio between the age and the half-mass relaxation time, as shown in Fig.~\ref{fig:alpha distribution}b. Specifically, $\alpha$ is consistent with the Salpeter index in young and intermediate-age clusters, while it starts to decrease for clusters older than $\sim 500$\,Myr. The same trend is also found by Ebrahimi and collaborators \citep[see e.g. Figure of]{ebrahimi2022a}. \\
To test the effect introduced by considering different stellar mass ranges in the determination of the cluster MFs, we adopted the following approach. Briefly, we simulated 1000 CMDs of clusters with the same age, distance, and metallicity as the observed clusters, and following a Salpeter MF. We the added errors in color and magnitude similar to the observed ones, and finally re-computed the MF of each simulation by following the same procedure and adopting the same mass interval adopted for the real clusters. Moreover, we randomly varied the observed stellar mass ranges by adding random Gaussian fluctuations. We find that the recovered MF slopes are consistent with the Salpeter input one to the level of 0.01. We can therefore safely assume that the bias introduced by analyzing different stellar mass intervals does not significantly affect our results. Moreover, in the case of a double-slope input MF, the recovered single slope is characterized by a larger spread with respect to the single-slope case. Specifically, in the case of a 1.35/2.35 input double slope MF, we find that the overall spread of the recovered single slope is of order 0.4.

\begin{figure*}[ht!]
    \centering
    \includegraphics[width=0.9\textwidth, trim={0cm 0cm 0cm 0cm}, clip]{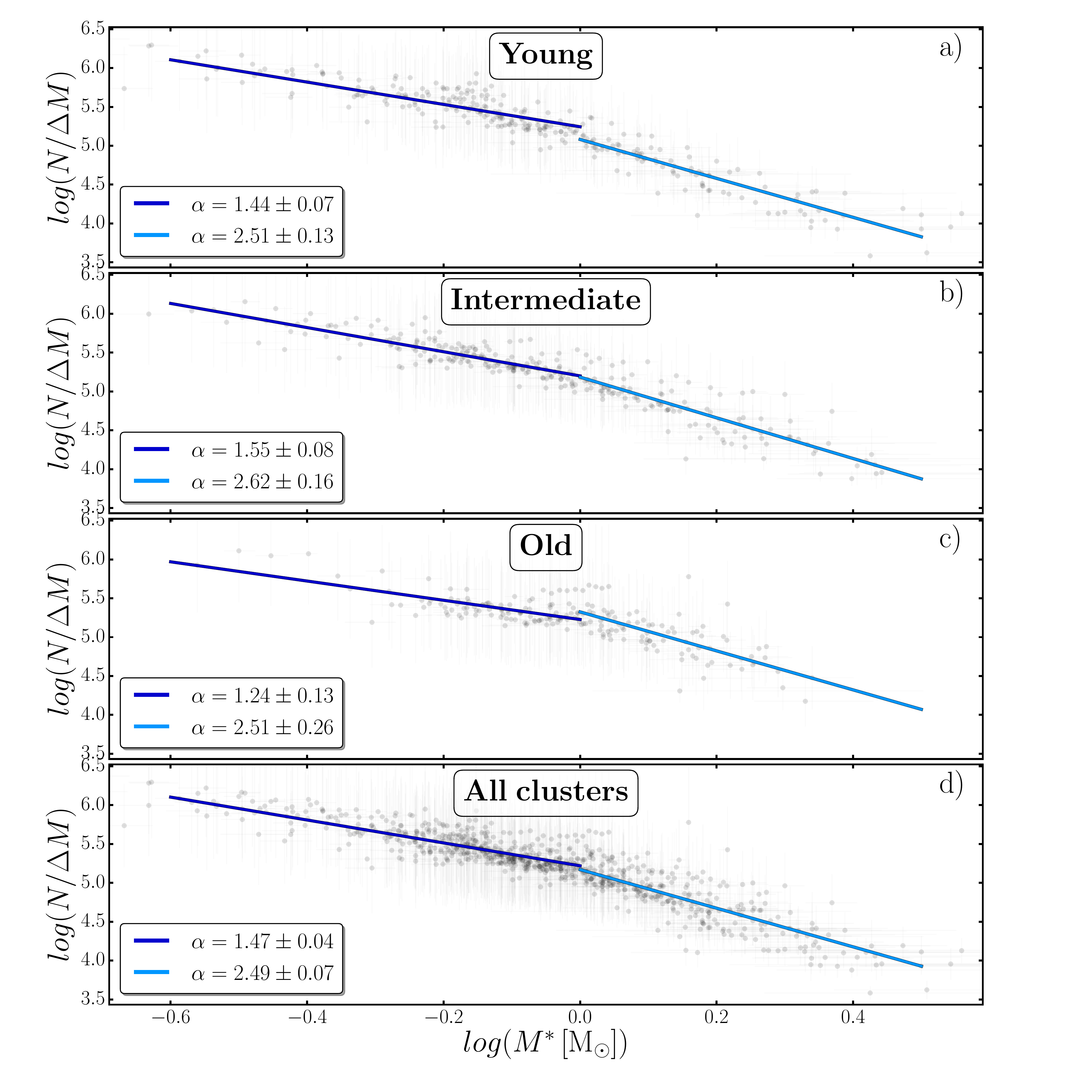}
    \caption{A common mass function. \textit{Panels (a)-(c).} Normalized stacked MFs for young ($log(Age/t_{\rm rh})\leq 0.2\,$), intermediate ($0.2<log(Age/t_{\rm rh})<0.75$), and old clusters ($log(Age/t_{\rm rh})\geq 0.75$). The normalization choice is described in Section~\ref{subsec:mass}. Blue and azure solid lines represent the straight line best-fit of the low-mass ($M\leq 1M_{\odot}$) and high-mass ($M> 1M_{\odot}$) stars, respectively. The derived slope is indicated in the bottom-left legends. \textit{Panel (d).} Same as panels (a)-(c), but for the whole sample of analyzed clusters.}
    \label{fig:stacked MF}
\end{figure*}

To compare the MFs of clusters with different ages, we normalized each MF to the number of stars with $-0.2\leq log(M/M_{\odot}) \leq 0.05$, which is the mass
interval common to all the analyzed clusters (see Figure~\ref{fig:alpha distribution}c).  
Figure~\ref{fig:stacked MF} shows the stacked MF for dynamically young, intermediate, and old clusters (panels (a), (b), and (c), respectively), and for all clusters (panel (d)). Specifically, we define dynamically young clusters as those with $log(age/t_{\rm rh})< 0.2$, intermediate those with $0.2 \leq log(age/t_{\rm rh})<0.75,$ and old as those with $log(age/t_{\rm rh})\geq 0.75$. The adopted boundaries are defined so that approximately the same number of clusters fall into each group.
 
These figures corroborate the evidence for a change in the MF slope around $1\,M_{\odot}$. 
We derived the curves that provide the least-squares best fit of the stacked MFs by using the stars more massive and less massive than $1\,M_{\odot}$ separately. The results are provided by the azure and blue straight lines shown in Figure~\ref{fig:stacked MF}.
The analysis shows that the MFs of the dynamically young, intermediate, and old clusters follow approximately the same pattern, with the slopes of low- and high-mass stars being consistent within $1\sigma$. 
Finally, we repeated the procedure including all clusters, regardless of their age. The result is shown in Fig.~\ref{fig:stacked MF}d. Remarkably, we find that the general trend is, again, well approximated by a double power-law MF, with slopes consistent with those found in \citet{sollima2019a} for $\sim 120,000$ stars in the solar neighborhood; that is, 1.34 and 2.41/2.68 in the sub- and supersolar regimes, respectively.

\section{The binary fraction}
\label{sec:bin}

The unresolved binary systems composed of two MS stars (MS-MS binaries) exhibit redder colors and brighter magnitudes than both stellar components. Hence, they lie on the bright and red side of the MS fiducial line in most CMDs constructed with the Gaia filters.
The exact location of MS-MS binaries in the CMD depends on the mass of the primary stars $(M_{\rm prim})$ and on the ratio between the mass of the secondary and primary star, $q=M_{\rm sec}/M_{\rm prim}$. 
The equal-mass binary systems (i.e., $q=1$) form a sequence parallel to the MS but $\sim$0.75 magnitudes brighter. On the other hand, binaries with $q$ close to zero would approach the MS fiducial line. For intermediate mass-ratio values, we observe that the smaller the value of $q$, the closer the binary system will lie to the MS fiducial.

\begin{figure*}[ht!]
    \centering
    \includegraphics[width=0.3\textwidth, trim={0cm 0cm 0cm 0cm}, clip]{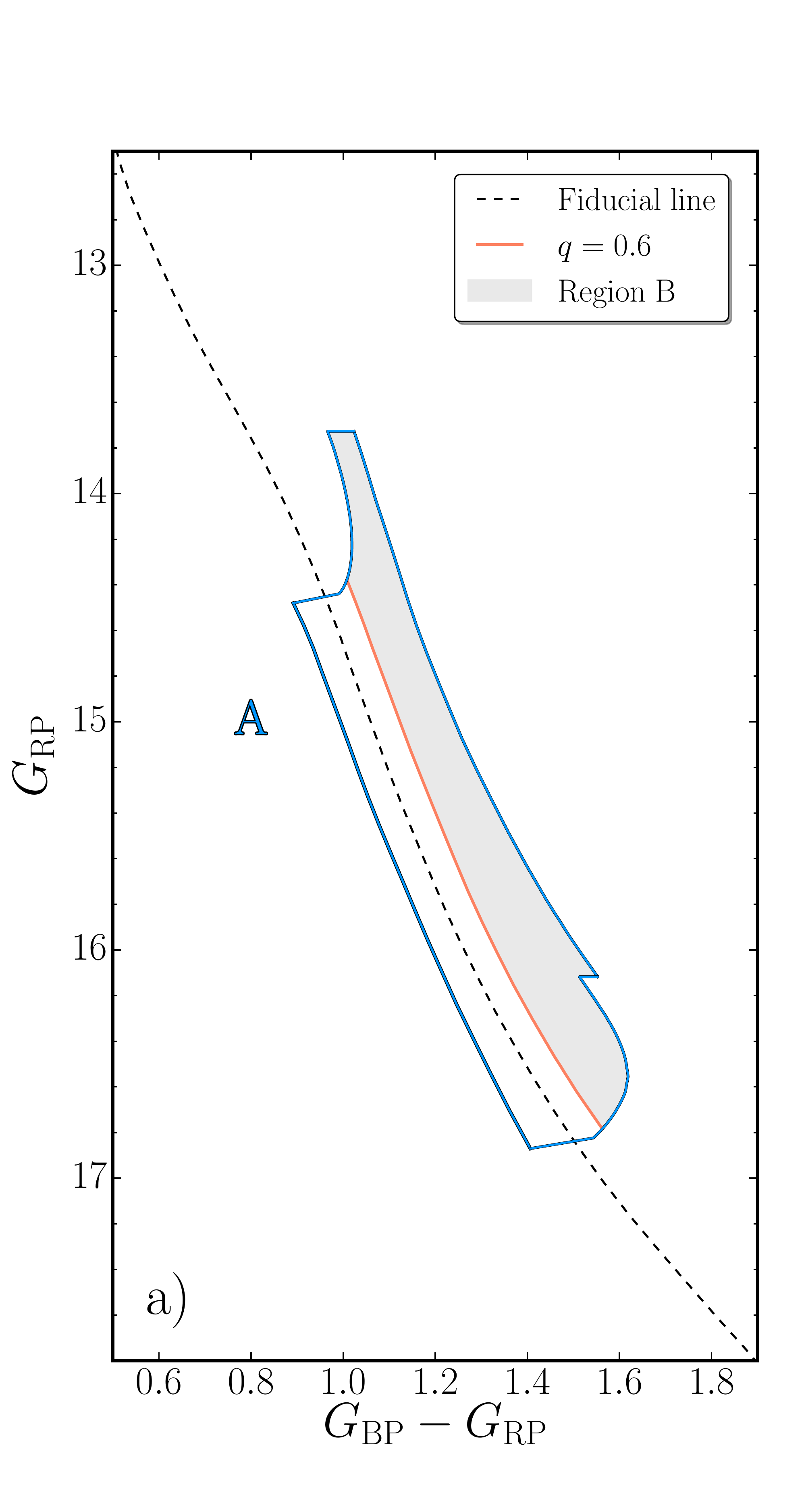}
    \includegraphics[width=0.3\textwidth, trim={0cm 0cm 0cm 0cm}, clip]{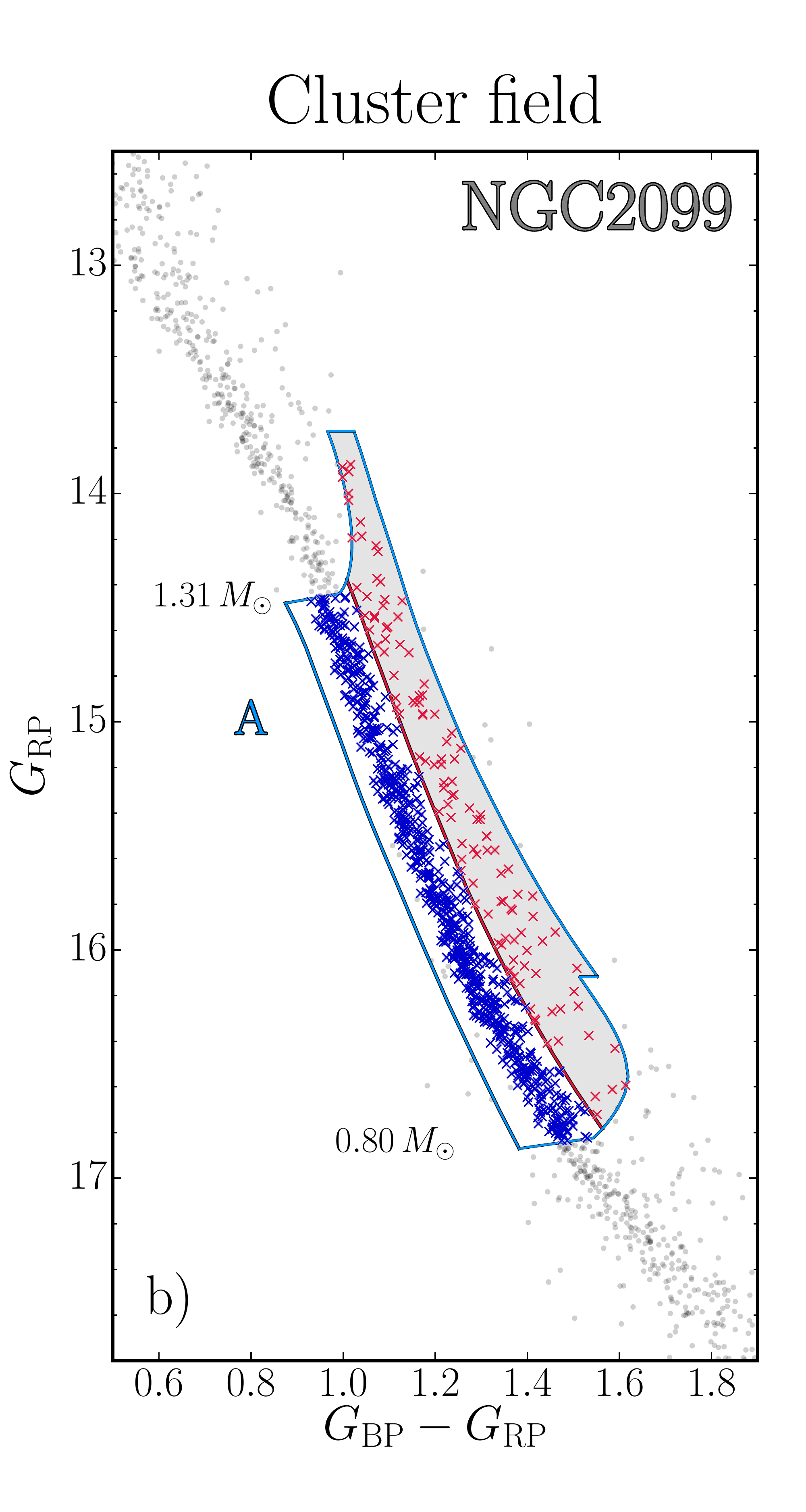}
    \includegraphics[width=0.3\textwidth, trim={0cm 0cm 0cm 0cm}, clip]{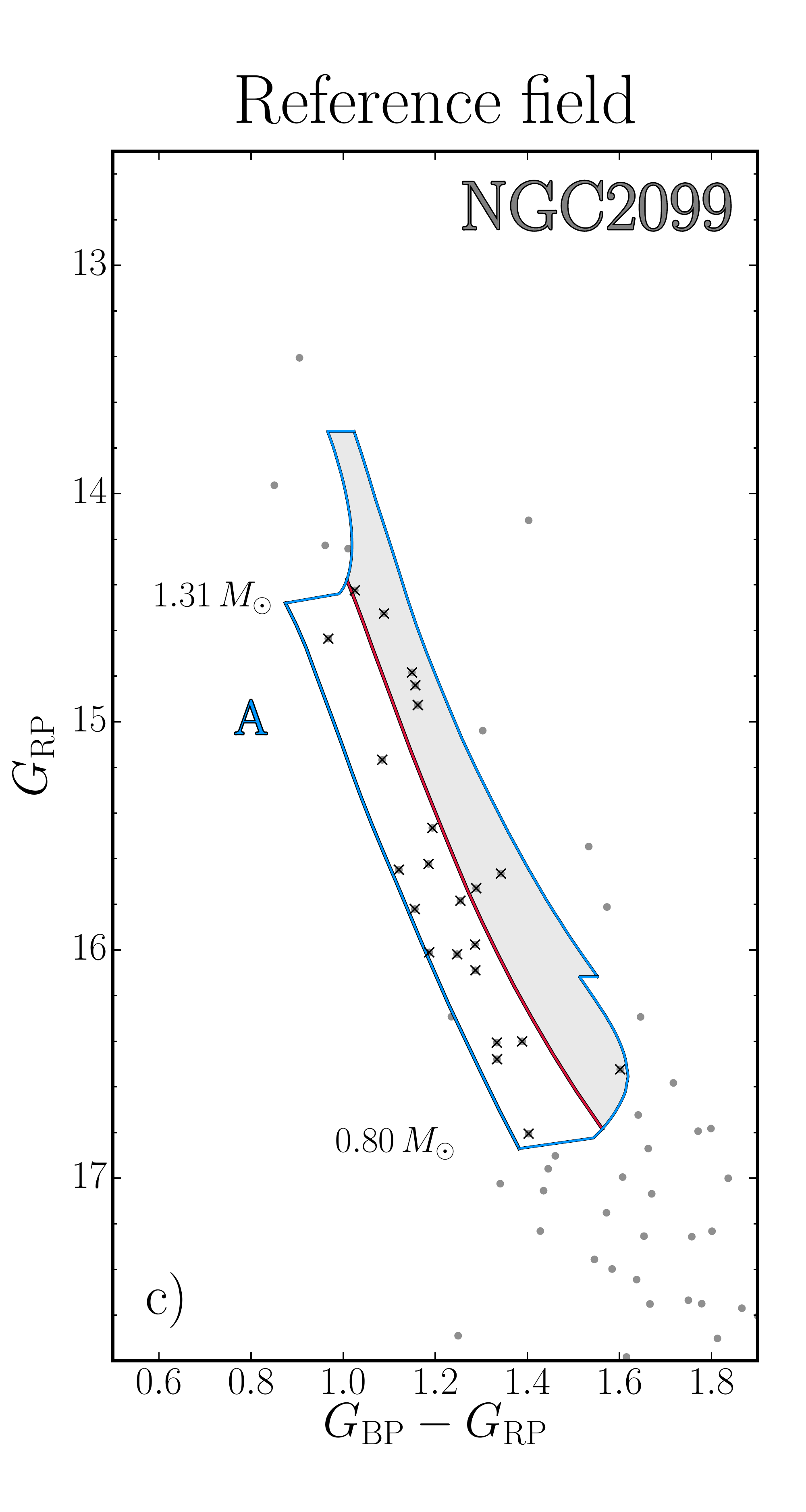}
    \caption{Binary stars selection. \textit{Panel (a):} Illustration of the binary-selection procedure. The region enclosed by the azure solid line, dubbed region A, contains all MS stars and binary stars with a primary mass between 0.80 and 1.31 $M_{\odot}$. The gray-shaded region, i.e., region B, defines the locus of binary-star candidates with a mass ratio greater than 0.6, respectively. Finally, the dashed black line and red solid lines represent the fiducial line of MS stars and binary stars with mass ratio $q=0.6$, as indicated in the top-right legend. \textit{Panel (b)}: MS and binary stars identified in the cluster field of NGC\,2099, marked with red and blue crosses, respectively. \textit{Panel (c):} Same as panel (b) but for stars located in the reference field, well outside the cluster radius. Stars that fall into region A are marked with black crosses. }
    \label{fig:bin selection}
\end{figure*}

To estimate the fraction of unresolved binary systems, we follow the approach of \citet{milone2012a} and \citet{cordoni2018a}, which is illustrated in Figure~\ref{fig:bin selection} for NGC\,2099.
For each cluster, we limited the analysis to binaries with a mass ratio greater than a fixed threshold value, $q_{\rm lim}$, ranging from 0.6 to 0.7 according to the overall quality of the cluster CMD. Indeed, binaries with small mass-ratio values, for example $q<q_{\rm lim}$, lie too close to the MS and are almost indistinguishable from single stars. 

First, we defined region A, which is enclosed by the azure line in Fig.~\ref{fig:bin selection}, and includes all single and binary cluster stars with $G_{\rm RP}^{\rm bright}\leq G_{\rm RP}\leq G_{\rm RP}^{\rm faint}$. Region B is instead defined as the portion of region A containing only candidate MS-MS binary stars with $q>q_{\rm lim}$, and is colored in gray in Fig~\ref{fig:bin selection}(a-c).

To define these two regions, we first derived the fiducial line of MS stars. To do this, we divided the MS into magnitude bins and computed the median and the 68.27$^{\rm th}$ percentiles of the color and magnitude distributions of the stars of each bin (hereafter $\sigma_{\rm GBP-GRP}$ and  $\sigma_{\rm GRP}$). The MS fiducial line (dotted line in Fig.~\ref{fig:bin selection}a) is derived by linearly interpolating these median points.

Specifically, the boundaries of region A, which is enclosed by the azure line in Figure~\ref{fig:bin selection}, are defined as follows: (i) the blue border is obtained by shifting the fiducial line four times $\sigma_{GBP-GRP}$ to the left; (ii) the bright and fainter boundaries are the loci of binary systems where the primary stars have magnitudes $G_{\rm RP}^{\rm bright}$ and $G_{\rm RP}^{\rm faint}$, respectively, and mass ratios ranging from 0 to 1; (iii) the reddest border is defined by shifting the fiducial line of equal mass binaries to the red by four times  $\sigma_{G_{\rm BP}-G_{\rm RP}}$.

Region B, shaded in gray in Figure~\ref{fig:bin selection}, is the portion of Region A that includes binaries with $q>q_{\rm lim}$. It is located on the red side of the fiducial line of binary stars with $q=q_{\rm lim}$. To avoid significant contamination from single stars, we have chosen the value of $q_{\rm lim}$ of each cluster in such a way that the fiducial line of binaries with $q=q_{\rm lim}$ is redder than the MS fiducial shifted by four times $\sigma_{G_{\rm BP}-G_{\rm RP}}$. We adopted $q_{\rm lim}=0.6$ or $q_{\rm lim}=0.7$ depending on the uncertainties of the photometry of each cluster.

Stars that fall into region B in NGC\,2099 are colored in red in Figure~\ref{fig:bin selection}b, while the remaining stars that lie in region A are marked with blue crosses.
To account for the residual contamination from field stars in the cluster CMD, we derived the numbers of field stars with cluster-like proper motions, and the parallaxes in the CMD of stars in the reference field normalized to the same area as the cluster field (see Figure~\ref{fig:bin selection}c for NGC\,2099).
The fraction of binaries is estimated as:

\begin{equation}
    f_{\rm bin}^{q\geq q_{\rm lim}} = \frac{N_{\rm cl}^{\rm B}-N_{\rm fl}^{\rm B}}{N_{\rm cl}^{\rm A}-N_{\rm fl}^{\rm A}}
    \label{eqn:bin fraction}
,\end{equation}

where the subscripts ``${\rm cl}$'' and ``${\rm fl}$'' refer to the cluster and the reference field, respectively.
We used a similar formula to calculate the fraction of binaries within the core. 

\begin{figure*}[ht!]
    \centering
    \includegraphics[width=\textwidth, trim={0cm 0cm 0cm 0cm}, clip]{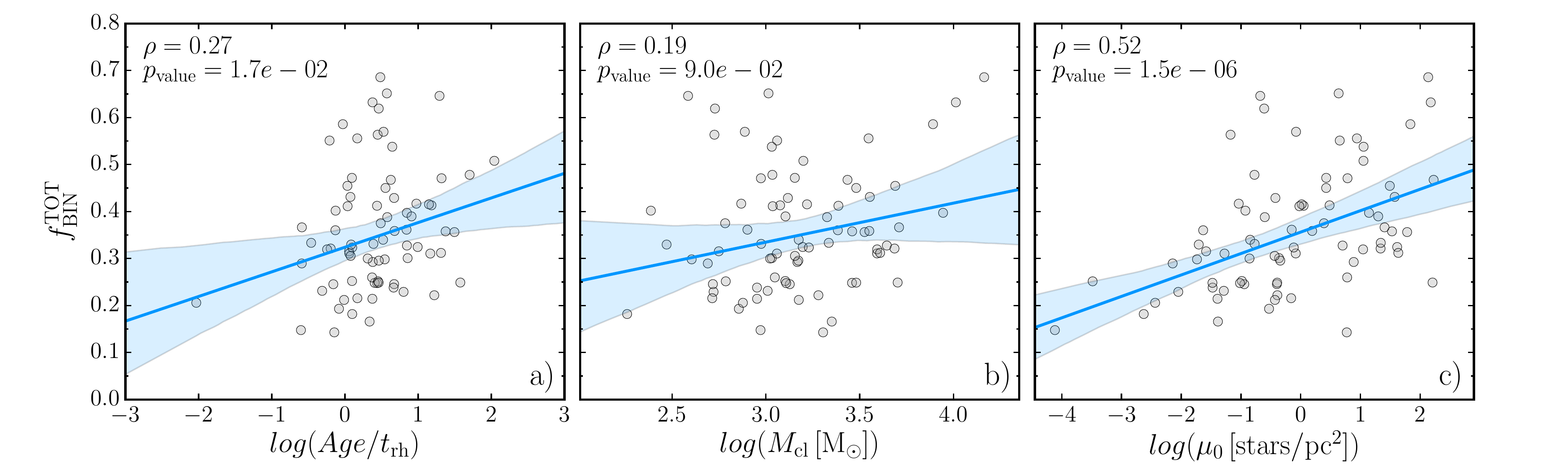}
    \includegraphics[width=\textwidth, trim={0cm 0cm 0cm 0cm}, clip]{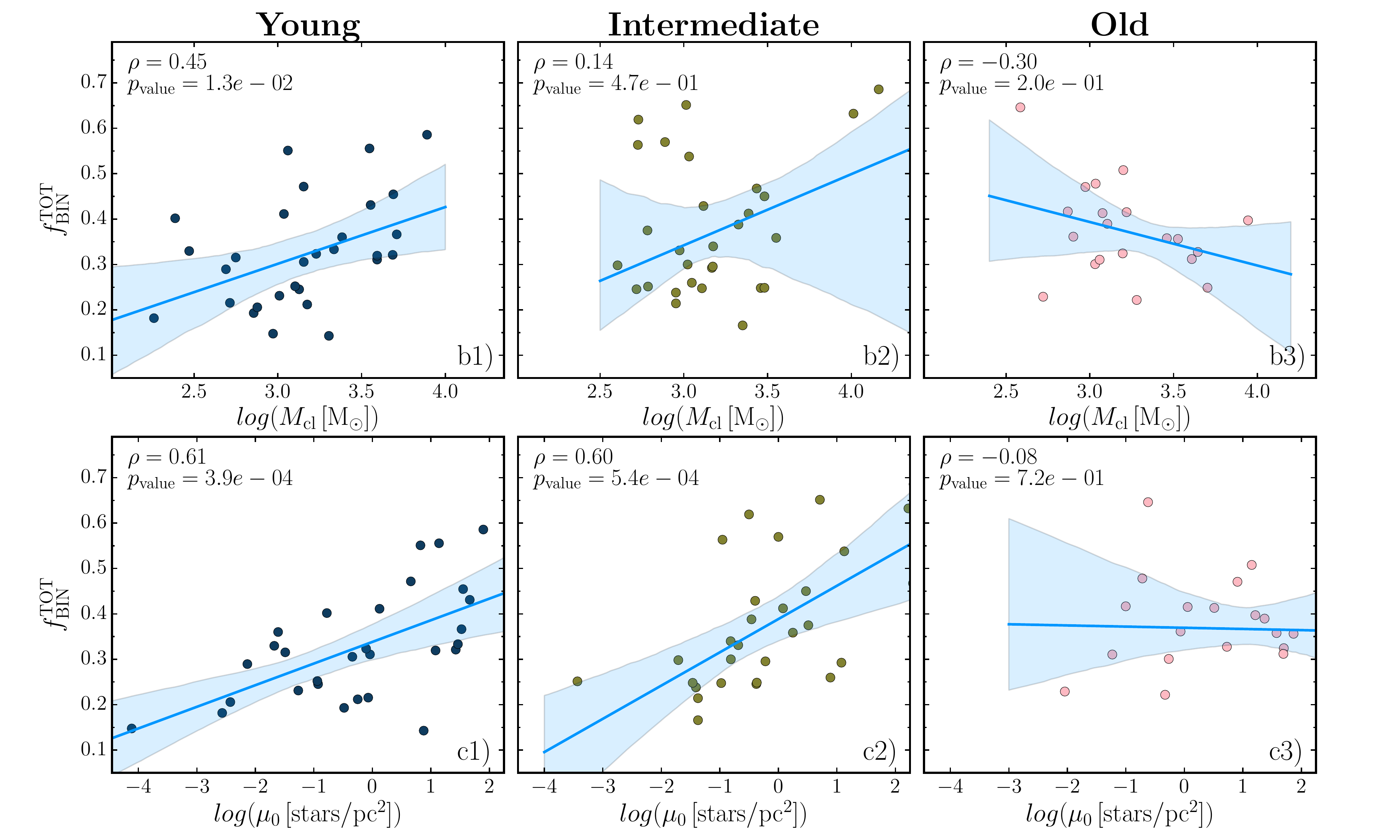}
    \caption{Relation between binary fractions and host clusters. \textit{Panels (a)-(c).} Total binary fraction as a function of cluster dynamical age, i.e., age normalized over the half-mass relaxation time, total cluster mass,  and central density in $\rm stars/pc^2$. Azure lines and shaded regions indicate the straight line that best fits the data and its $1\sigma$ dispersion. The values of Spearman's rank coefficient $(\rho$) and the corresponding $p$-values are indicated in the top-left corner of each plot. \textit{Panels (b1)-(b3).} Same as panel (b), but with clusters grouped into three dynamical age groups, defined as young, intermediate, and old clusters. \textit{Panels (c1)-(c3)}. Same as panel (c), for the three groups of clusters.}
    \label{fig:results}
\end{figure*}

The fractions of binaries range from less than 0.2 to $\sim$0.7.  As shown in Fig.~\ref{fig:results}, there is no evidence for significant correlations between the fractions of binaries and or between cluster age and mass. However, Fig.~\ref{fig:results}c reveals a mild correlation between the total binary fraction and the central density derived in Sect.~\ref{sub:radial}. Moreover, to account for the different distances of the analyzed clusters, we converted the central density from $\rm stars/arcmin^2$ to $stars/pc^2$. In this section, we discuss only the relevant correlations between the total and core binary fractions and the physical parameters of the 
clusters; however, the plot  in the Appendix  shows all the correlations (see e.g. Fig.~\ref{fig:all corr 0}-\ref{fig:all corr 1}). Remarkably, while there is no correlation between the total binary fraction and the age or the dynamical age of a cluster ($\rho =0.20, 0.27$ respectively, bottom row of Fig.~\ref{fig:all corr 0}), there are hints of a mild correlation between the core binary fraction and cluster age and dynamical age ($\rho =0.54, 0.49$ respectively, bottom row of Fig.~\ref{fig:all corr 1}). We investigate these relationships in the following sections. \\
To further explore possible connections between binary fractions and the physical parameters of the clusters, we divided the clusters into the three ``age'' groups defined in Fig.~\ref{fig:stacked MF} and repeat the analysis. Remarkably, the total binary fraction mildly correlates with cluster mass for the dynamically young cluster (Fig.~\ref{fig:results}b1), with $log(age/t_{\rm rh})\leq0.2$, while it weakly anti-correlates in old clusters (Fig.~\ref{fig:results}b3). Concerning the central stellar density, Figs.~\ref{fig:results}c1-c3 reveal that the binary fraction is linked with the central density in young and intermediate age clusters, while these two quantities seem to be uncorrelated in older clusters. 

\subsection{Binaries and blue stragglers}
\begin{figure}
    \centering
    \includegraphics[width=0.3\textwidth]{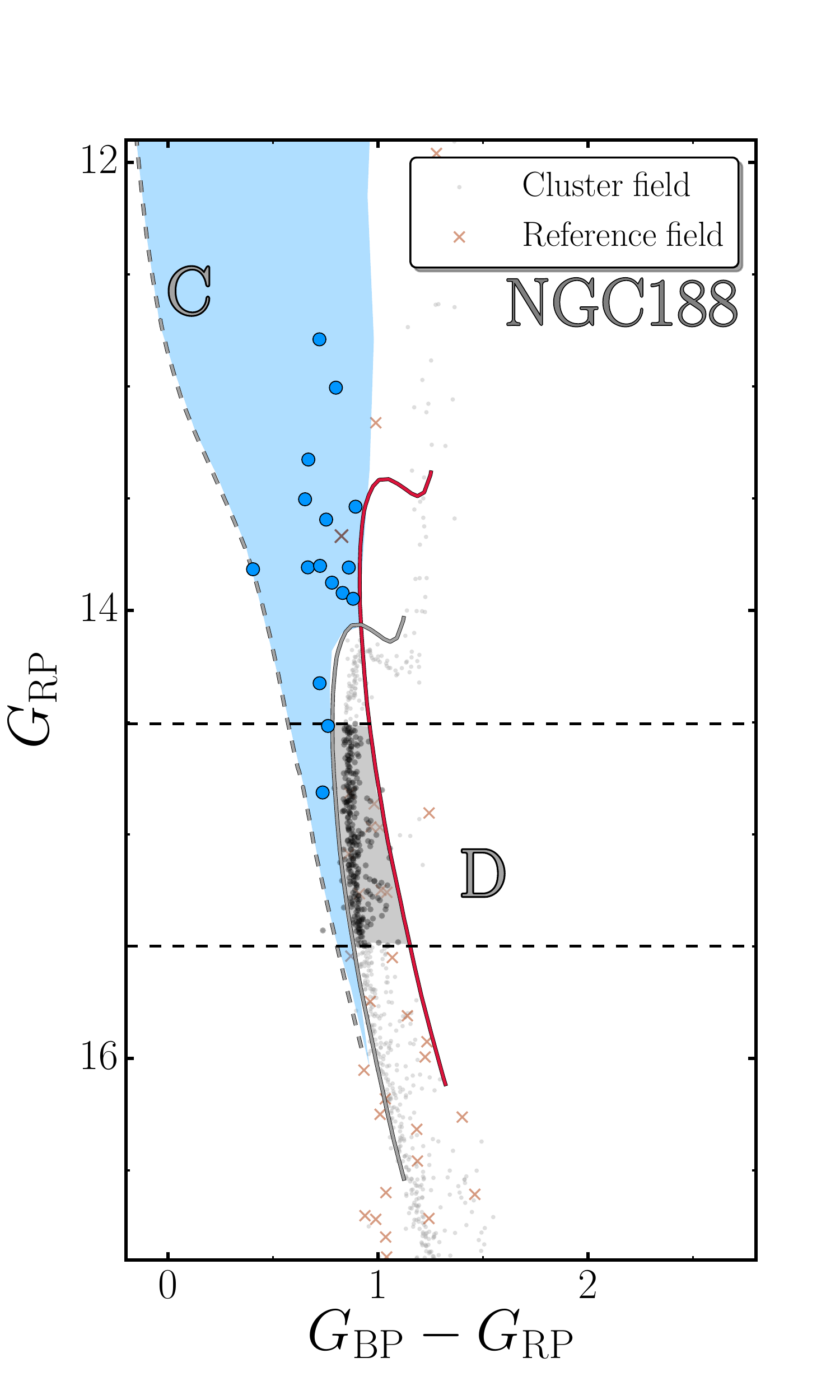}
    \caption{
     Procedure used to estimate the fraction of BSSs in NGC\,188. Cluster and reference-field stars are marked with gray circles and brown crosses, respectively. The gray dashed line represents the ZAMS. 
     The gray solid line is the fiducial line of the MS and the SGB shifted to the blue by four times the observational color error ($\sigma$). The fiducial line of MS-MS and SGB-SGB binaries, shifted by four times $\sigma$ to the red is represented by the red line.
     Region C of the CMD is indicated as an azure-shaded area, while region D, marked with the gray shade, is enclosed by gray and red solid lines and by the two horizontal black dashed lines. 
     Cluster field stars in regions C and D of the CMD are
     marked with azure solid and black-shaded circles, respectively.}
    \label{fig:bss selection}
\end{figure}

To investigate the relations between binaries and blue stragglers (BSSs), we followed the procedure illustrated in Fig.~\ref{fig:bss selection} for NGC\,188.
We first defined the regions C and D in the CMD of stars in the cluster field. Region C, indicated by the azure shade, is populated by candidate BSSs. The left border is defined by the zero-age MS (ZAMS) isochrone (dashed line), while the right boundary is provided by the combination of the MS and Sub Giant Branc (SGB) fiducial line blue-shifted by four times the color error (grey line), and the fiducial line of equal-mass binaries, shown in red.

Region D, shaded in grey, is mostly populated by MS stars within 1 magnitude of the MS turn-off, that is, with $G_{\rm RP}$ in the range $(G_{\rm RP}^{\rm Toff}, G_{\rm RP}^{\rm Toff}+1)$.
The fraction of BSSs is estimated as
\begin{equation}
   f_{\rm BSS}=\frac{N_{\rm cl}^{\rm C}-N_{\rm ref}^{\rm C}} {N_{\rm cl}^{\rm D}-N_{\rm ref}^{\rm D}}
,\end{equation}
where $N_{\rm cl}^{\rm C}$ and $N_{\rm cl}^{\rm D}$ are the numbers of stars in the regions C and D, respectively, in the cluster-field CMD, while $N_{\rm ref}^{\rm C}$ and $N_{\rm ref}^{\rm D}$ are the corresponding number of stars in the reference field.

\begin{figure*}[ht!]
    \centering
    \includegraphics[width=0.99\textwidth, trim={0cm 0cm 0cm 0cm}, clip]{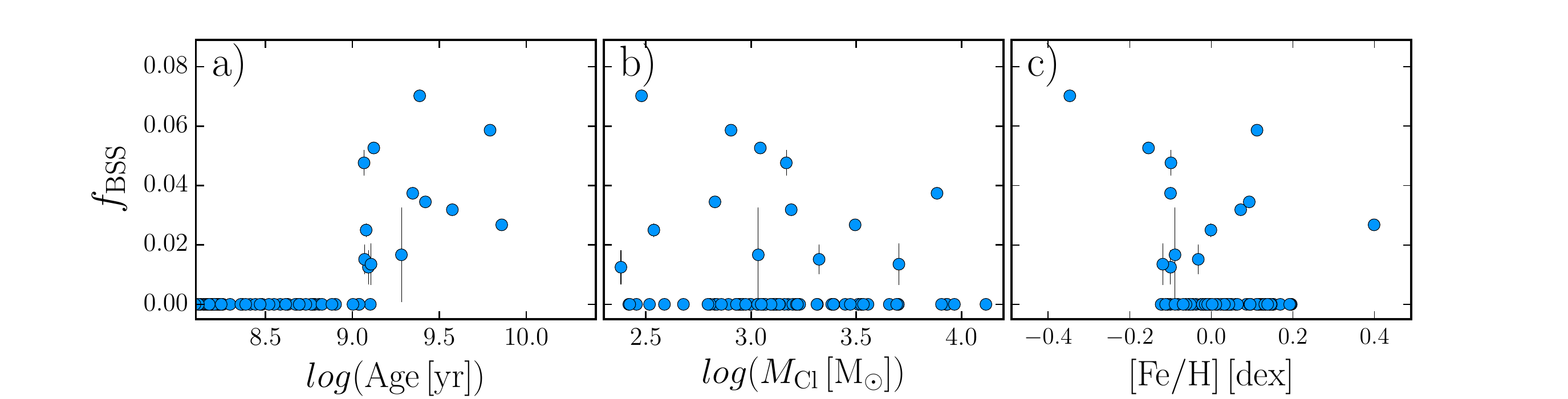}
    \caption{Blue stragglers properties. Fraction of BSSs against cluster age (panel a), cluster mass (panel b), and cluster metallicity.}
    \label{fig:bss corr}
\end{figure*}

We confirm the previous conclusion that clusters younger than $\sim$300 Myr do not host BSSs \citet{vaidya2022a, rain2021a}. 
As shown in Figure~\ref{fig:bss corr}, the fraction of BSSs shows no evidence of a correlation with cluster mass or metallicity, but there is a mild correlation with cluster age \citep[see also][]{rain2021a}.

\begin{figure}[ht!]
    \centering
    \includegraphics[width=0.45\textwidth, trim={0cm 0cm 0cm 0cm}, clip]{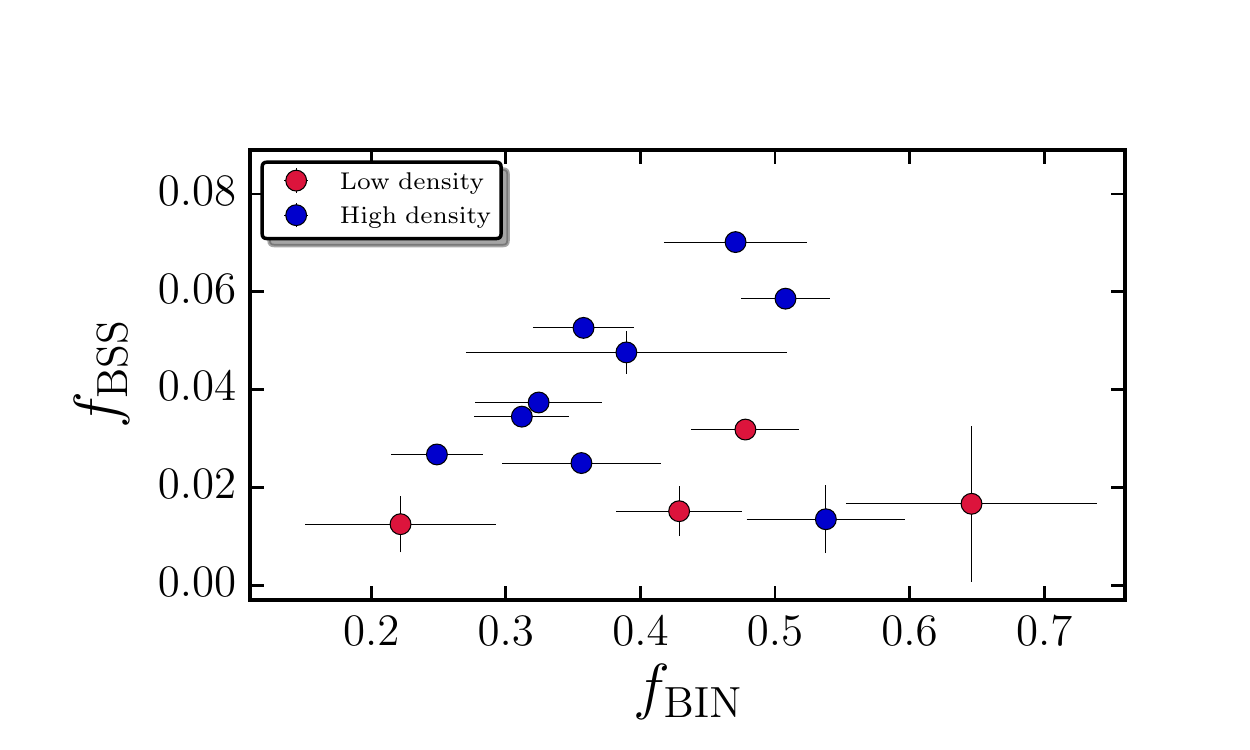}
    \caption{Fraction of blue stragglers vs.\,binary fraction. Red colors indicate the clusters with central density $\mu_{0}<10 \;\rm stars/pc^2$, whereas the remaining clusters are colored blue.}
    \label{fig:bss bin}
\end{figure}

 Figure~\ref{fig:bss bin} compares the fraction of BSSs and the binary fraction and shows no evidence of a significant correlation. Nevertheless, when we consider two groups of clusters with different central density values, expressed in $\rm stars/pc^2$ (see Section~\ref{subsec:radius}). We note that low-density  clusters ($\mu_{0}<1\rm stars/pc^2$) define a sequence that extends toward large values of $f_{\rm bin}$ but   small fractions of BSSs, whereas the remaining clusters (blue dots) have, on average, larger values of $f_{\rm BSS}$ and define a steeper sequence in the $f_{\rm BSS}$ vs.\,$f_{\rm bin}$ plane. 
 The high-density cluster Tombaugh\,1 is a possible exception.

\subsection{Mass-ratio distribution}\label{sub:massratio}
\begin{figure}
    \centering
    \includegraphics[width=0.49\textwidth, trim={9.2cm 4.5cm 13.0cm 3cm}, clip]{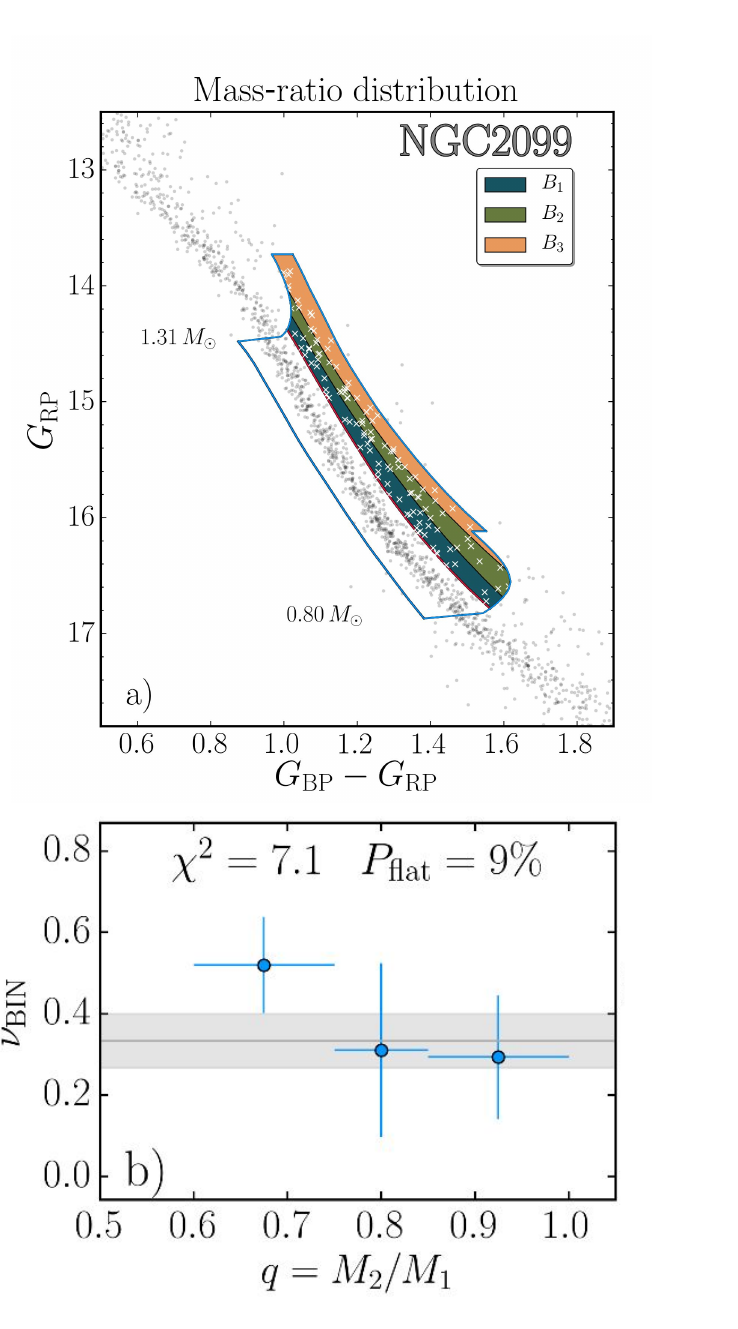}
    \caption{Binary stars mass-ratio as a function of mass-ratio in NGC\,2099.
    Reproduction of the CMD of NGC 2099 plotted in Fig.~\ref{fig:bin selection}b (panel a). The shaded regions (B1, B2, and B3) of the CMD comprise binaries with different mass ratios.  The candidate binaries are marked with white crosses. Panel (b) shows the mass-ratio distribution of binaries in NGC 2099. The $\chi^2$ and the probability that the observed binary fractions come from a flat mass-ratio distribution are quoted in panel (b).}
    \label{fig:ngc2099 qdist}
\end{figure}

To investigate binaries with different mass ratios, we adapted the  method of \citet{milone2012a} to open clusters. Briefly, we divided region B of the CMD into three regions, B1, B2, and B3 as shown in Fig.\,\ref{fig:ngc2099 qdist}a for NGC\,2099, where the candidate binaries are marked with white crosses. Each region encloses binaries with different mass ratios and the values of $q$ corresponding to its left and right boundaries are derived with the criterion that each region covers similar areas in the CMD. Each subregion comprises binaries in a different mass-ratio interval $\Delta q_{\rm i}$. To compare the binary fractions inferred for the three mass-ratio intervals, we estimated the normalized binary fraction: $\nu_{\rm i}=f_{\rm bin, i}/\Delta_{\rm i}$, where the index i ranges from 1 to 3, and the $f_{\rm bin, i}$ is the fraction of binaries in the regions B1, B2, and B3 of the CMD.
 
Results are shown in Fig.~\ref{fig:ngc2099 qdist}b for NGC 2099, where we plot $\nu_{\rm i}$ as a function of $q$. The horizontal gray line and shaded region indicate the total normalized binary fraction with $q>q_{\rm lim}$ and its uncertainty, respectively.
 
We performed a $\chi^2$
test to investigate whether or not the binaries with $q>q_{\rm lim}$ of each cluster are consistent with a flat mass-ratio distribution. We first computed the $\chi^2$ between the observed  $\nu_{\rm i}$  values and the flat mass-ratio distribution corresponding to the horizontal gray line of Fig.\,\ref{fig:ngc2099 qdist}a, $\chi^2_{\rm OBS}$. We then generated 1000 CMDs with the same number of stars and the same fraction of binaries with $q>q_{\rm lim}$ as the observed one, but with a flat mass-ratio distribution. We calculate the normalized binary fractions, $\mu_{\rm i}$, corresponding to each simulation and derived the corresponding $\chi^2$ values with respect to the flat mass-ratio distribution, in close analogy with what we did for the observed CMD, $\chi^2_{\rm SIM}$.

The probability that the observed distribution is consistent with a flat trend is estimated as the fraction of simulations with $\chi^2_{\rm SIM} > \chi^2_{\rm OBS}$ and corresponds to one in the case of a flat distribution. 
As an example, the observed mass-ratio distribution of binaries in NGC\,2099 has a $9\%$ probability of being flat. 

Results for all clusters are shown in Fig.\,\ref{fig:massratio_dist_all}, while the values of the $\chi^2$ and $P_{\rm flat}$ are provided in Table~\ref{tab:tab2}. We find that in 66 out of 78 clusters, the mass-ratio distribution has a probability higher than 10\% of being flat. 

\begin{figure}[ht!]
    \centering
    \includegraphics[width=0.45\textwidth, trim={2cm 0cm 27.0cm 0cm}, clip]{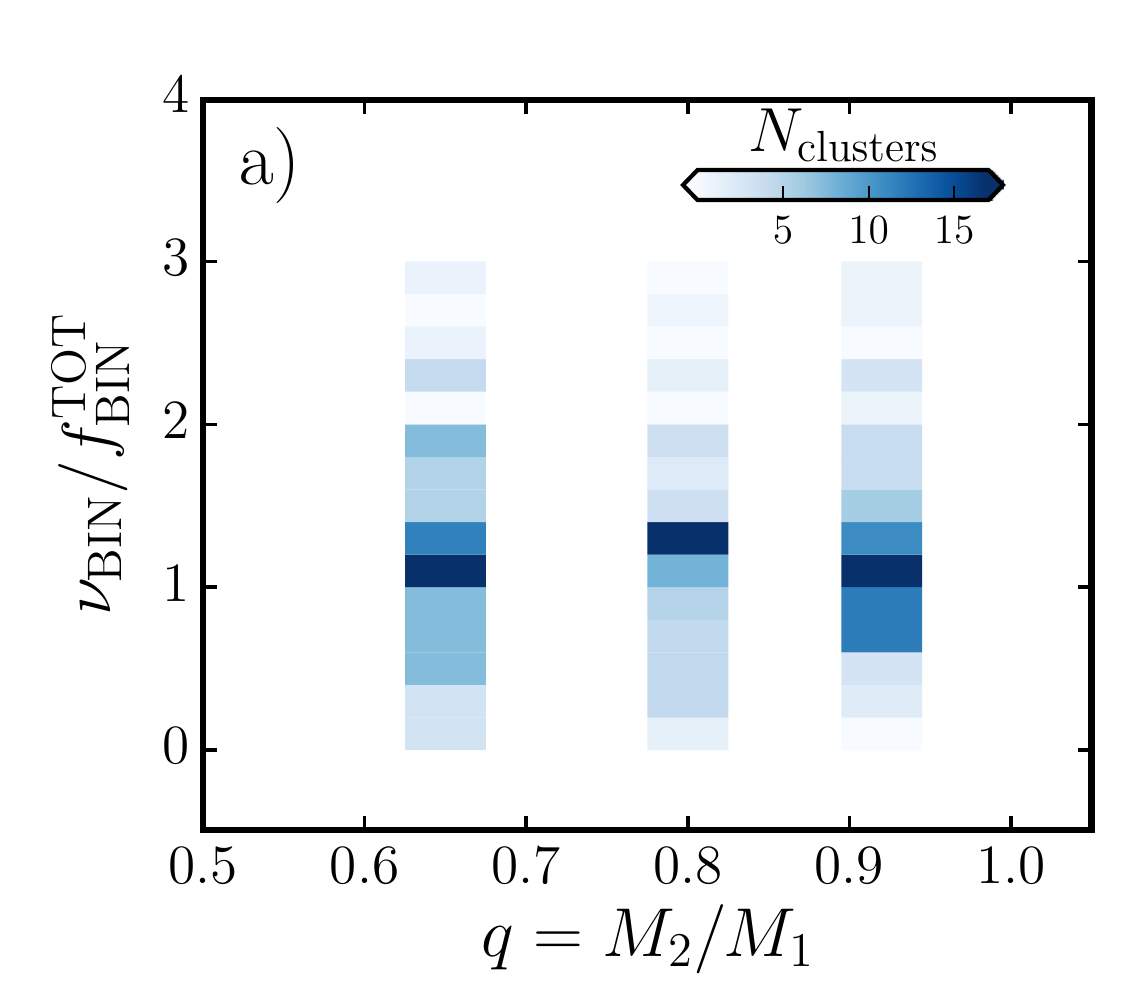}
    \caption{Binary fraction mass-ratio distribution. The binary fraction in each mass-ratio bin has been normalized over the width of the bin, in $q$, and the total fraction of binaries extrapolated assuming a flat mass ratio distribution. Each cell is color coded according to the number of observations in that particular bin, as specified in the top right color bar.}
    \label{fig:FbinVSq}
\end{figure}

To further investigate the mass-ratio distribution of binary systems, we combine the results from all clusters as shown in Fig.~\ref{fig:FbinVSq}. To do this, we divided the normalized binary fractions by the total fraction of binaries. The latter is estimated as $f_{BIN}^{\rm TOT}=f_{\rm BIN}^{q>q_{\rm lim}}/(1-q_{\rm lim})$.
We adopted a 2D-histogram representation to show the distribution of the normalized fraction of binary stars in each mass-ratio bin. The color code is indicated by the top-right color bar. 
When results from all clusters are combined together, we find a flat mass-ratio distribution for binary systems with $q\geq 0.6$, as can be inferred by the position of the darkest cells in Fig.~\ref{fig:FbinVSq}. 

\subsection{Binary fraction as a function of the  mass of the primary star}\label{sub:mass}

\begin{figure}
    \centering
    \includegraphics[width=8cm, trim={8.5cm 4.5cm 13.0cm 0cm}, clip]{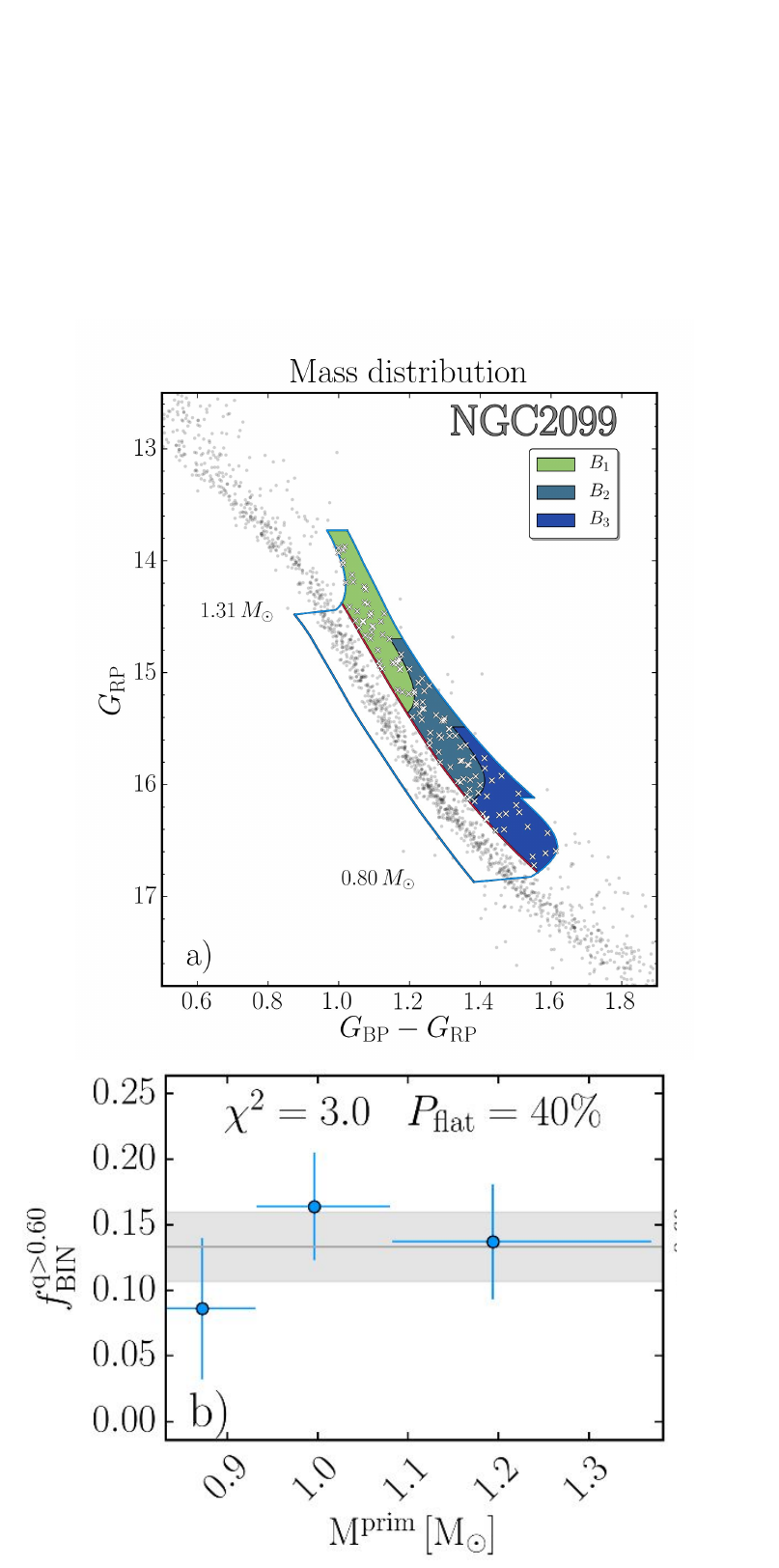}
     \caption{Binary stars as a function of primary stars $M_{\rm prim}$ in NGC\,2099. The adopted regions are colored differently, as defined in the main text. The mass distribution of candidate binary stars with $q\geq \_{\rm lim}$ is shown in panel \textit{(b)}.}
    \label{fig:ngc2099 mdist}
\end{figure}

To determine the fraction of binary systems as a function of primary star mass, we divided region A of the CMD into three different subregions, as indicated in Fig.~\ref{fig:ngc2099 mdist}, each containing approximately the same number of stars. We computed the fraction of binary stars in each region by means of Equation~\ref{eqn:bin fraction}. 
The resulting distribution is shown in panel (b2), where the horizontal gray line and gray shaded region indicate the binary fraction of NGC\,2099. Visual inspection of the mass distribution reveals a weakly increasing or flat trend, as indicated by the $P_{\rm flat}$ value defined in the previous section. Specifically, the value of the $\chi^2$ and its probability prevent us from making further conclusions. 
Overall, we find that 72 out of 78 clusters are consistent with a flat mass distribution of binary stars, with $P_{\rm flat}>10\%$. 

\begin{figure}[ht!]
    \centering
    \includegraphics[width=0.45\textwidth, trim={13.5cm 0cm 15.5cm 0cm}, clip]{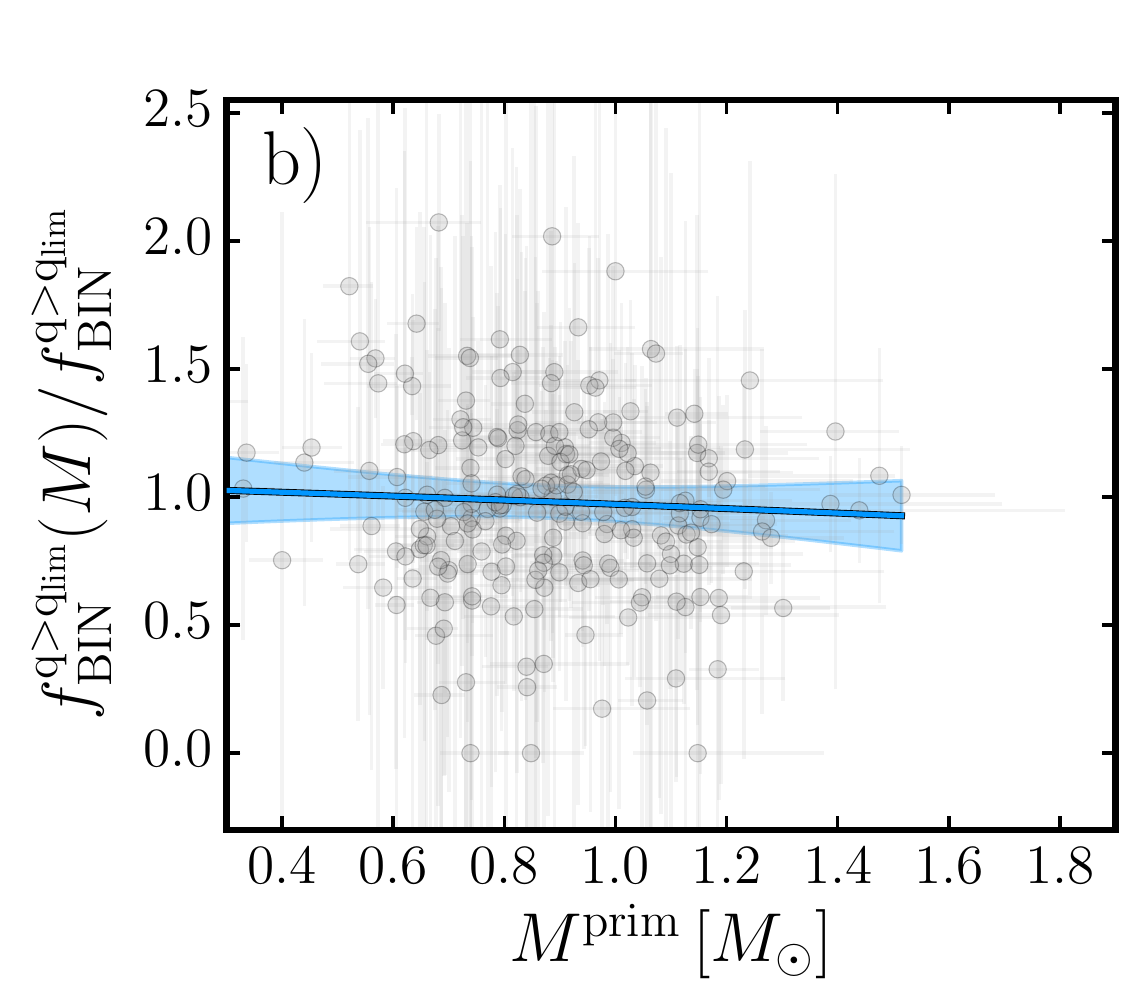}
    \caption{Binary fraction primary star distribut. The azure line and the shaded region represent the median trend and the 1$\sigma$ confidence interval.}
    \label{fig:FbinVSmass}
\end{figure}

\subsection{Radial distribution}\label{sub:radial}
To investigate the radial distribution of binary systems, we divided the field of view into three circular regions containing approximately the same number of stars.
For each radial bin, we estimated the binary fraction with the procedure of Section~\ref{sec:bin}.

The results are illustrated in Fig.~\ref{fig:rad dist all}, where we plot the fraction of binaries as a function of the radial distance from the cluster center.
We find that 15 clusters out of the 78 analyzed here exhibit a low probability of having a flat radial distribution, that is,
$P_{\rm flat}<5\%$.

\begin{figure*}[ht!]
    \centering
    \includegraphics[width=0.99\textwidth, trim={0cm 0cm 0cm 0cm}, clip]{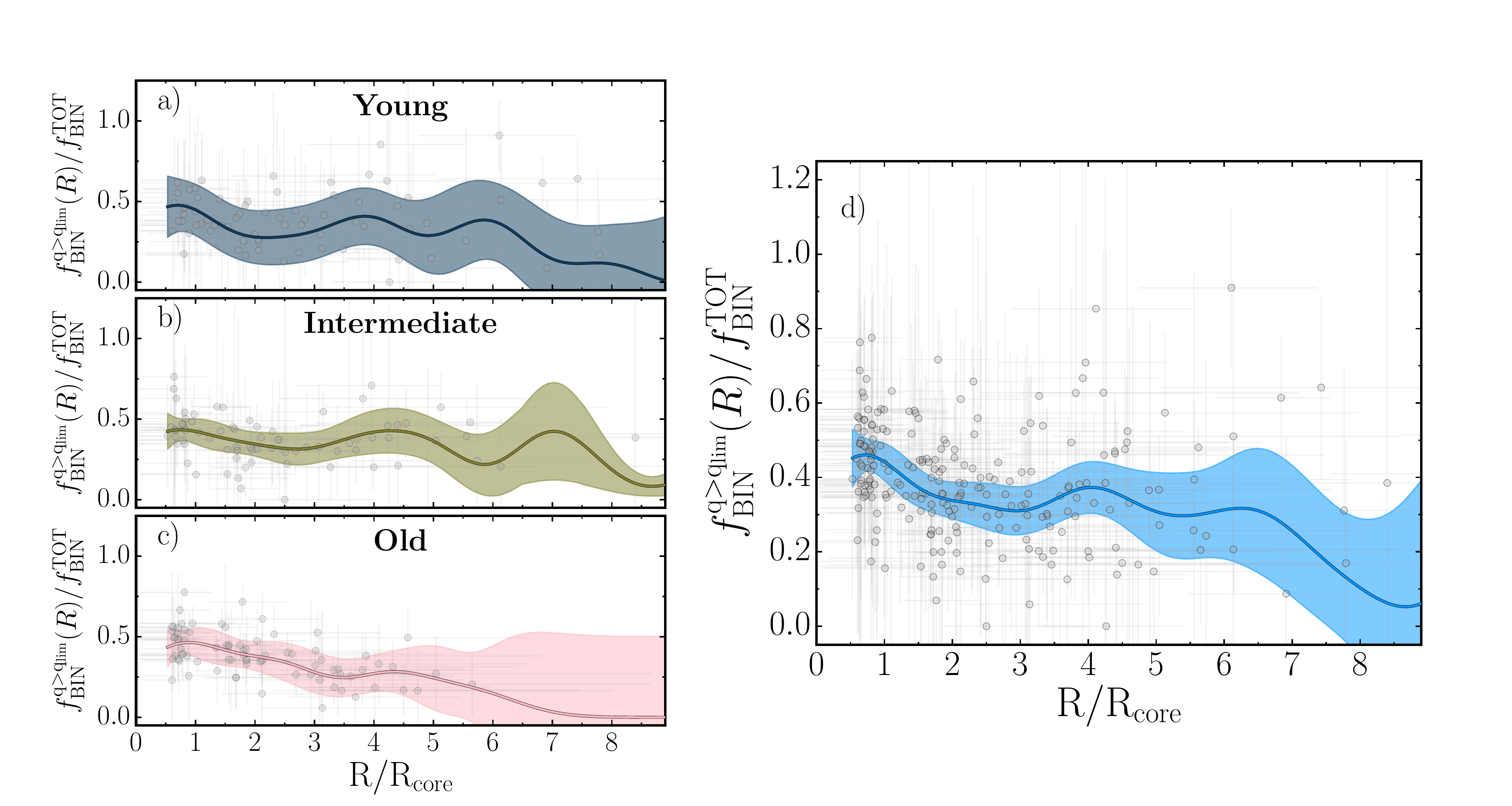}
    \caption{Binary stars radial distribution. Binary fraction with $q\geq q_{\rm lim}$ normalized to the total binary fraction, as a function radial distance from the cluster center in core radius units.  The azure lines mark the median trends, while the shaded areas correspond to the 1$\sigma$ dispersion. Panels (a) to (c) include clusters in the three different dynamical age intervals defined in the main text (see Subsection~\ref{subsec:mass}), while panel (d) refers to all clusters. Color coding and age bins are the same as in Figure~\ref{fig:alpha distribution}.}
    \label{fig:results_radial}
\end{figure*}

To further investigate the radial distribution of binary stars, we considered the three groups of clusters with different ages ---defined in Section~\ref{subsec:mass}--- separately.
For each age group, we plotted the fraction of binaries normalized to the total binary fraction against the radial distance from the cluster center normalized to the core radius.
As shown in Fig.~\ref{fig:results_radial}(a--c), we find that the binaries of both young and intermediate-age clusters (i.e., clusters younger than $\sim 800 \,\rm Myr$) exhibit a nearly flat radial distribution, with some hints of two peaks corresponding to the cluster center and to a radius of about four times the core radius.
The binaries of the old clusters are more centrally concentrated than single stars. The binary fraction is maximum in the core and declines by a factor of two at the radial distance of four core radii. Such behavior is qualitatively consistent with the mild correlations between core binary fractions and cluster age. 
For completeness, we show in Fig.~\ref{fig:results_radial}d the radial distribution of binaries for the entire sample of studied clusters. 
 
\subsection{Comparison with Galactic globular clusters}
\label{subsec:gcs}
In this subsection, we compare our findings on Galactic open clusters with those of \citet{milone2012a} and \citet{milone2016a} for Galactic globular clusters (GCs). 
As the works by Milone and collaborators are limited to the central regions of GCs, we only compare the binaries in the core regions of globular and open clusters. Moreover, to minimize the effects of low statics on our conclusions, we restrict the investigation to the open clusters with more than 500 stars.

\begin{figure*}[ht!]
    \centering
    \includegraphics[width=0.99\textwidth, trim={0cm 0cm 0cm 0cm}, clip]{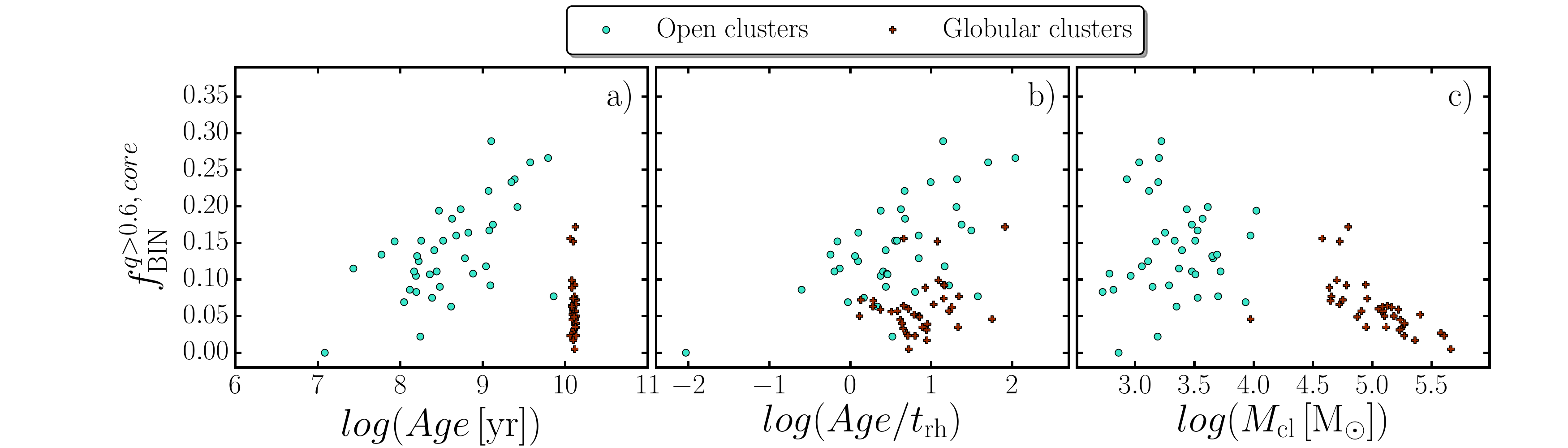}
    \caption{Comparison between OCs and GCs. Total fraction of binaries in the core of open clusters (aqua circles) and GCs (red crosses) as a function of  the logarithm of cluster age (panel a), dynamical age (panel c), and cluster mass (panel d).
    The GC binary fractions are taken from \citet{milone2012a, milone2016a}, while GC masses are from \citet{baumgardt2018a}.
    }
    \label{fig:ocgc}
\end{figure*}

Results are shown in Figure~\ref{fig:ocgc}, where we plot the total fraction of binaries in the core of Galactic open clusters (aqua-blue circles) and in GCs (red crosses) as a function of the logarithm of cluster age (panel a), dynamical age (panel b), and cluster mass (panel c), as derived in \citet{baumgardt2018a}.

Intriguingly, Fig.~\ref{fig:ocgc}(a) reveals that the observed core binary fractions with $q>0.6$ of OCs is larger than that of GCs. However, panel (b) of the same figure may suggest that some GCs, namely NGC\,6652, NGC\,6838, and Ruprecht\,106, characterized by a higher core binary fraction, are consistent with the trend defined by OCs. However, we observe that both GCs and OCs exhibit a qualitatively similar increasing pattern in the $f_{bin}$ vs. $log(age/t_{\rm rh})$ plane. On the other hand, we observe opposite relations between core binary fractions and cluster mass. Indeed, OCs exhibit a correlation, while GCs clearly define an anti-correlation. Possible exceptions may be represented once again, by NGC\,6652, NGC\,6838, Rup\,106, and NGC\,6835, which seem to follow the relation defined by OCs. However, a more detailed analysis is required to draw firm conclusions.   

While the fraction of binaries in the GC core anti-correlates with the mass of the host cluster, there is no evidence of a similar trend between the fraction of binaries in the core of open clusters and their masses.

\section{Summary and conclusion}
\label{sec:results}
We used the Gaia DR3 data to investigate the population of unresolved binaries along the MS of 78 Galactic open clusters. To this end, we first carefully selected the probable cluster members based on their parallaxes and proper motions. Moreover, we corrected the photometry for effects of differential reddening when needed and accounted for residual field star contamination.  We also exploited the isochrones from the Padova database \citep[][]{marigo2017a} that, according to \citet[][]{dias2021a}, provide the best match with the observed CMDs.

For each cluster, we measured the fraction of binaries with mass ratios larger than a fixed threshold, q$>$q$_{\rm lim}$, where q$_{\rm lim}$ ranges from 0.6 to 0.7, depending on the photometric errors. We also extrapolated the total binary fraction by assuming a flat mass-ratio distribution. We derived the binary fractions in the entire cluster field of view and in the region within the cluster core.
In addition, we investigated the fraction of binaries in three different mass-ratio bins and in different intervals of primary-star mass. Finally, we analyzed the radial distribution of binaries.
 
 To properly investigate the relations between binary stars and the host star cluster, we derived various cluster properties:
 \begin{itemize}
     \item   We estimated the EFF profile that provides the best fit of the observed cluster-density profile. We therefore provide the central density and the core radius  for each cluster.  
     \item  We derived the MF of each cluster and constrained the MF slope, $\alpha_{\rm MF}$, using a single power-law function. We find that the value of $\alpha_{\rm MF}$ depends on cluster age and ranges from $\sim$2.5 in $\sim$100 to 200 Myr-old clusters to less than 1.0 in clusters older than $\sim$2 Gyr. This finding is consistent with the recent conclusion by \citet[][]{ebrahimi2022a} based on 15 nearby open clusters. The MFs are used to estimate the total cluster mass.
     \item Visual inspection of the MFs of several clusters shows that stars in different mass intervals define MFs with different slopes. When we combine the MFs of all clusters together, we find that stars less massive than one solar mass are well fitted with a power-law function with $\alpha \sim 1.5$, whereas the MF of the other stars is consistent with  $\alpha \sim 2.5$.
     \item We derived the fraction of BSSs with respect to the MS stars in the luminosity interval between $G_{\rm RP}=0.0$ and 1.4 mag from the turn-off. We confirm that clusters younger than $\sim$300\,Myr show no evidence for BSSs \citep[see][]{rain2021a}. In the clusters with BSSs, we find a mild correlation between the fraction of BSSs and cluster age and no correction with cluster mass. 
 \end{itemize}
  
 Our main results on binaries can be summarized as follows:
 \begin{itemize}
     \item When we analyze the entire cluster field, the fraction of binaries ranges from $\sim 15\%$ in NGC\,2669 to more than 60\% in Haffner\,26. There is no significant correlation between the binary fractions and either cluster age or cluster mass. A similar conclusion can be extended to the binaries in the core. 
     Our results on open clusters differ from what was found in GCs, where there is a strong anti-correlation between the fraction of MS binaries in the core and the cluster mass \citep[][]{milone2012a, milone2016a}.
     \item There is no correlation between the fractions of binaries and BSSs. However, when we consider eight of nine star clusters with high central density ($\mu_{0}>1$star/pc$^{2}$), the fraction of BSSs correlates with the fraction of binaries.

     \item There is no evidence for relations between the fraction of binaries of most clusters and either the mass of the primary star or the mass ratio.
     \item The radial distribution of binaries significantly changes from one cluster to another and depends on cluster age. The binaries of the dynamically young clusters, that is,  those with $log(Age/t_{\rm rh})<0.2,$ show a nearly flat radial distribution, and the distribution of binaries in intermediate-age clusters, $0.2<log(Age/t_{rm rh}) < 0.75,$ exhibits hints of a double peak. On the contrary, the binaries of dynamically older clusters are more centrally concentrated than single stars, which is similar to what is observed in most GCs \citep[][]{milone2012a, milone2016a}. This trend is also corroborated by the correlations between core binary fraction and both cluster age and dynamical age, and the absence of any correlation between total binary fraction and either cluster age or dynamical age.
     Our findings are qualitatively consistent with the theoretical predictions by  \citet{geller2013} and \citet[][]{sollima2008a}. These authors used $N$-body simulations to model star clusters and binary evolution, and predict different radial distributions for binary stars of clusters with different ages. The fraction of binaries in young clusters (ages smaller than $\sim 300$ Myr) would increase towards the external regions, while the binaries stars clusters older than $\sim 1\,\rm Gyr$ would exhibit an opposite trend. The  clusters with intermediate ages would exhibit binaries with a double-peaked radial distribution.
     
     \item We find signs of a mild correlation between total binary fraction and central stellar density ($\rho=0.54$). Dividing clusters into three dynamical age groups, we find that this correlation is more evident in younger clusters, where the Spearman's rank correlation coefficient reaches 0.63 for young clusters and 0.59 for intermediate-age clusters. We find no sign of correlation in dynamically old clusters. Moreover, while there is no correlation between cluster mass and binary fraction when considering the whole sample of clusters, dynamically young clusters exhibit hints of a mild correlation, that is, $\rho=0.45$.  
 \end{itemize}

 The results discussed in this work provide important constraints to the properties and evolution of Galactic OCs. In particular, we derived the properties of stellar binary systems and blue stragglers and connect them to the structural and physical parameters of the host cluster.

\section*{Acknowledgments}
We thank the anonymous referee for various suggestions that im-
proved the quality of the manuscript.
This work has received funding from the European Research Council (ERC)
under the European Union’s Horizon 2020 research innovation programme
(Grant Agreement ERC-StG 2016, No 716082 ’GALFOR’, PI: Milone,
http://progetti.dfa.unipd.it/GALFOR) and from the European Union’s Horizon 2020 research and innovation programme under the Marie SklodowskaCurie Grant Agreement No. 101034319 and from the European Union –
NextGenerationEU, beneficiary: Ziliotto. APM, MT, and ED acknowledge
support from MIUR through the FARE project R164RM93XW SEMPLICE
(PI: Milone). APM and ED have been supported by MIUR under PRIN program 2017Z2HSMF (PI: Bedin). AFM acknowledges support from the project ``Understanding the formation of globular clusters with their multiple stellar generations''. EV acknowledges support from NSF grant AST-2009193. SJ acknowledges support from the NRF of Korea (2022R1A2C3002992, 2022R1A6A1A03053472)

\appendix

\section{Open cluster CMDs}
In the following, we show the differential reddening free  $G_{\rm RP}$ vs. $G_{\rm BP}-G_{\rm RP}$ CMDs of the 78 analyzed open clusters. The CMDs include only cluster members, and are sorted according to the age of the cluster, indicated in the top inset in gigayears. 
\begin{figure*}[ht!]
    \centering
    \includegraphics[width=0.85\textwidth, trim={1cm 2cm 1cm 0cm}, clip]{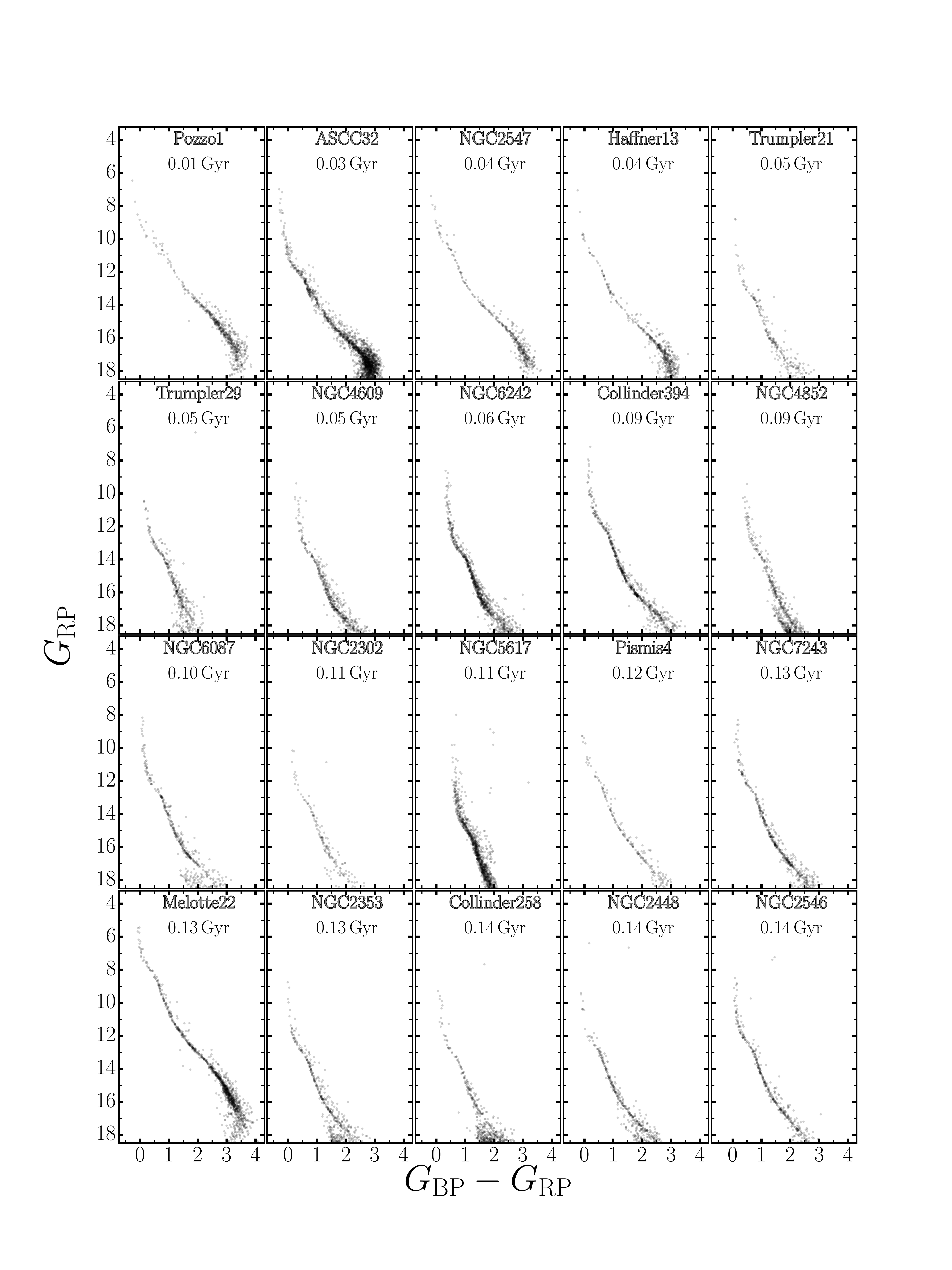}
    \caption{$G_{\rm RP}$ vs. $G_{\rm BP}-G_{\rm RP}$ CMDs corrected for  differential reddening whenever needed. The clusters are sorted according to age \citep[from][]{dias2021a}: Pozzo\,1, ASCC\,32, NGC\,2547, Haffner\,13, Trumpler\,21, Trumpler\,29, NGC\,4609, NGC\,6242, Collinder\,394, NGC\,4852, NGC\,6087, NGC\,2302, NGC\,5617, Pismis\,4, NGC\,7243, Melotte\,22, NGC\,2353, Collinder\,258, NGC\,2448, and NGC\,2546. }
    \label{fig:cmds1}
\end{figure*}

\begin{figure*}[ht!]
    \centering
    \includegraphics[width=0.87\textwidth, trim={1cm 2cm 1cm 0cm}, clip]{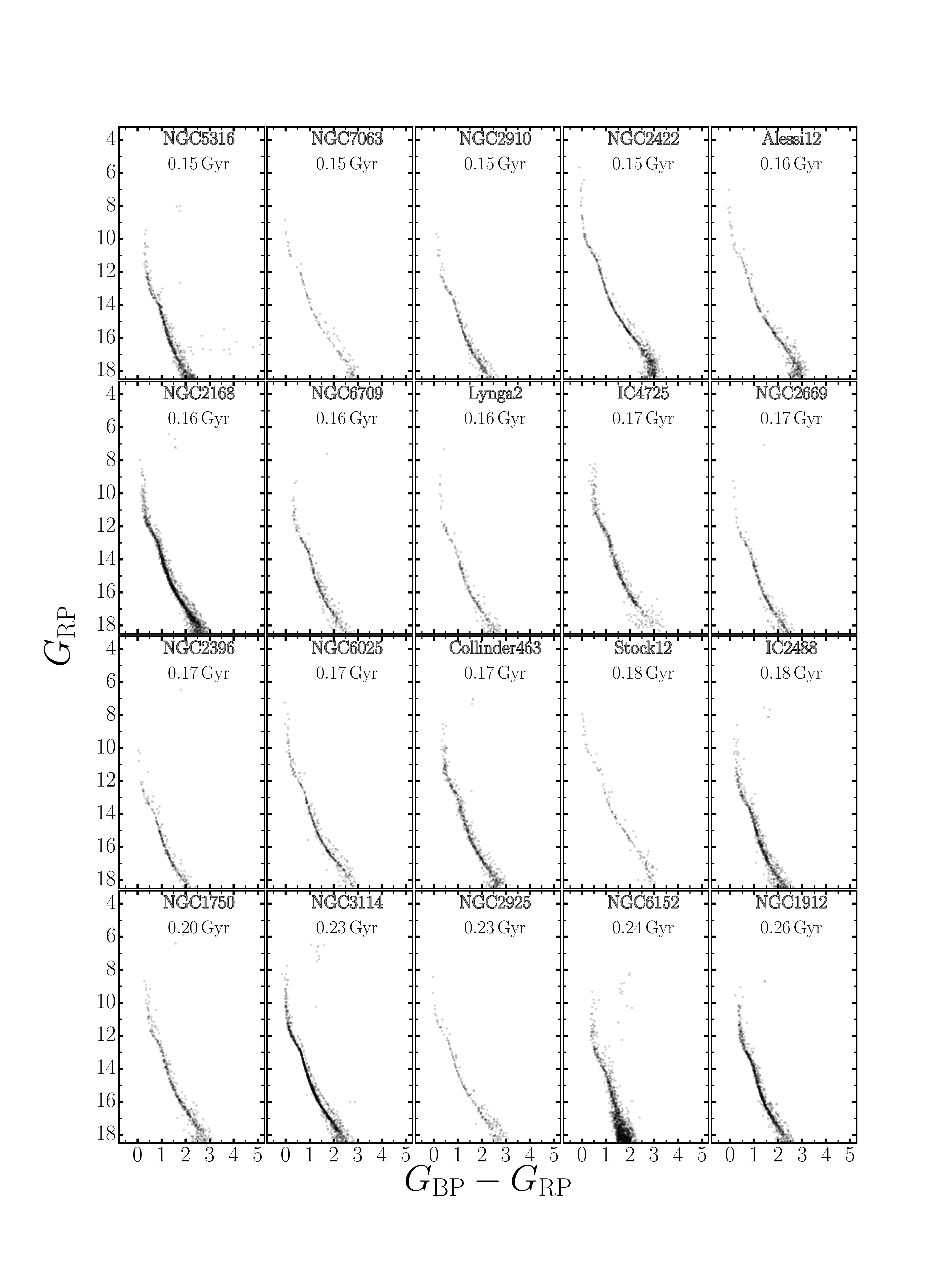}
    \caption{Same as Fig.~\ref{fig:cmds1}, but for the clusters NGC\,5316, NGC\,7063, NGC\,2910, NGC\,2422 , Alessi\,12, NGC\,5168, NGC6709, Lynga\,2, IC\,4725, NGC\,2669, NGC\,2396, NGC\,6025, Collinder\,463, Stock\,12, IC\,2488, NGC\,1750, NGC\,3114, NGC\,2925, NGC\,6152, and NGC\,1912.}
    \label{fig:cmds2}
\end{figure*}

\begin{figure*}[ht!]
    \centering
    \includegraphics[width=0.87\textwidth, trim={1cm 2cm 1cm 0cm}, clip]{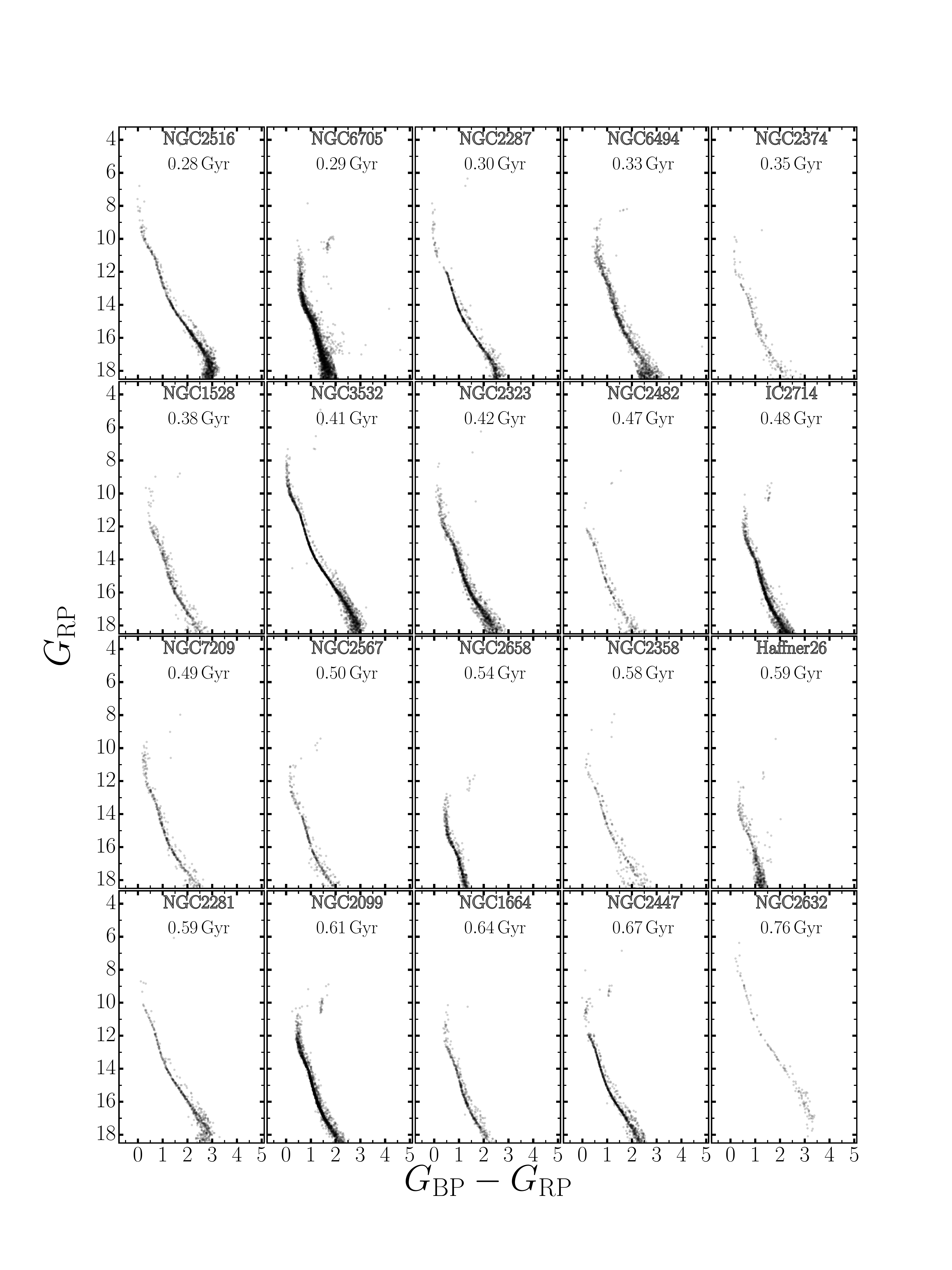}
    \caption{Same as Figure~\ref{fig:cmds1}, but for the clusters NGC\,2516, NGC\,6705, NGC\,2287, NGC\,6494, NGC\,2374, NGC\,1528, NGC\,3532, NGC\,2323, NGC\,2482, IC\,2714, NGC\,7209, NGC\,2567, NGC\,2658, NGC\,2358, Haffner\,26, NGC\,2281, NGC\,2099, NGC\,1664, NGC\,2447, and NGC\,2632. }
    \label{fig:cmds3}
\end{figure*}

\begin{figure*}[ht!]
    \centering
    \includegraphics[width=0.87\textwidth, trim={1cm 2cm 1cm 0cm}, clip]{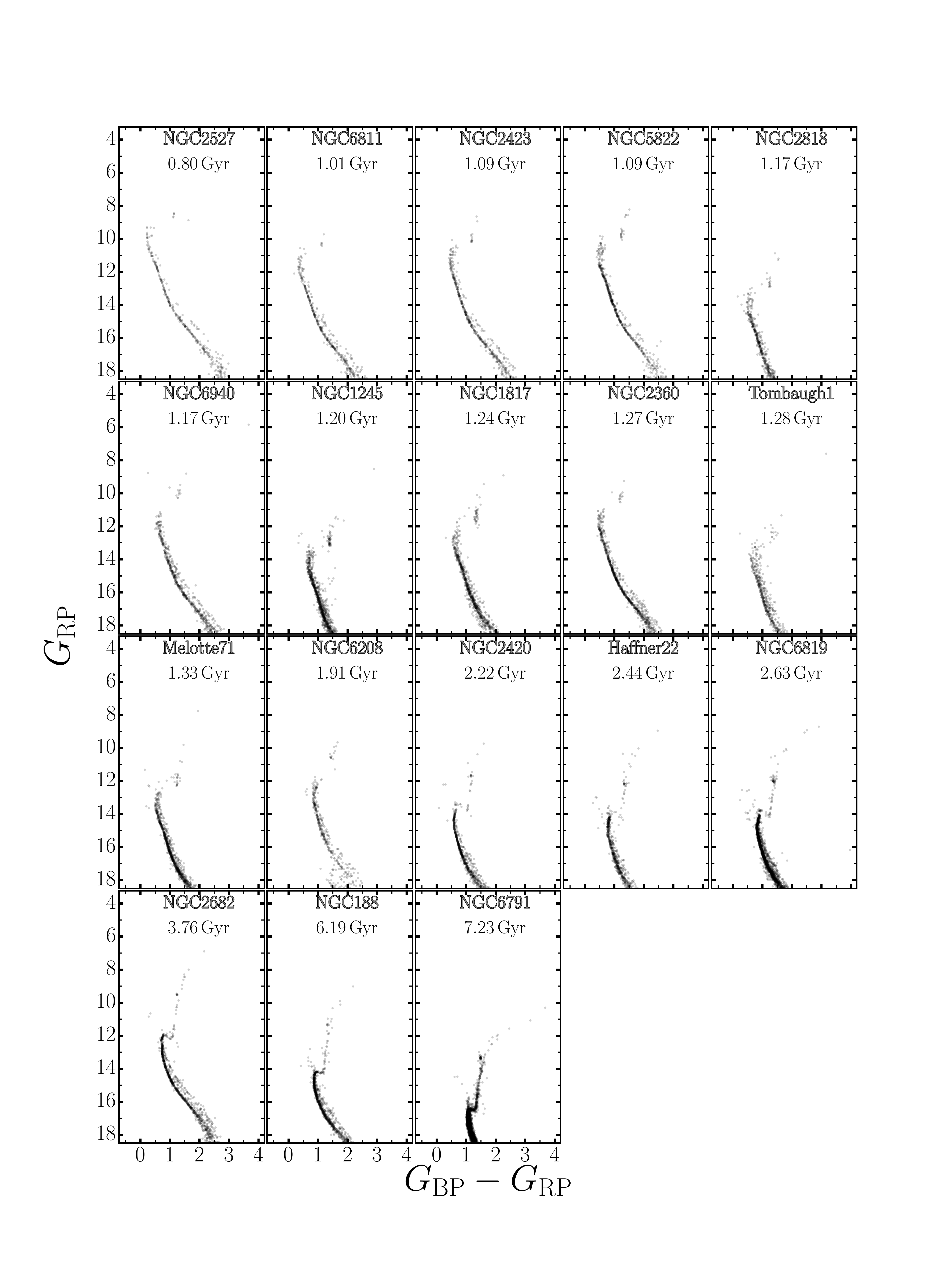}
    \caption{Same as Figure~\ref{fig:cmds1}, but for the clusters NGC\,2527, NGC\,6811, NGC\,2423, NGC\,5822, NGC\,2818, NGC\,6940, NGC\,1245, NGC\,1817, NGC\,2360, Tombaugh\,1, Melotte\,71, NGC\,6208, NGC\,2420, Haffner\,22, NGC\,6819, NGC\,2682, NGC\,188, and NGC\,6791. }
    \label{fig:cmds4}
\end{figure*}

\section{Individual binary-star distributions}
\begin{figure*}
    \centering
    \includegraphics[width=0.9\textwidth]{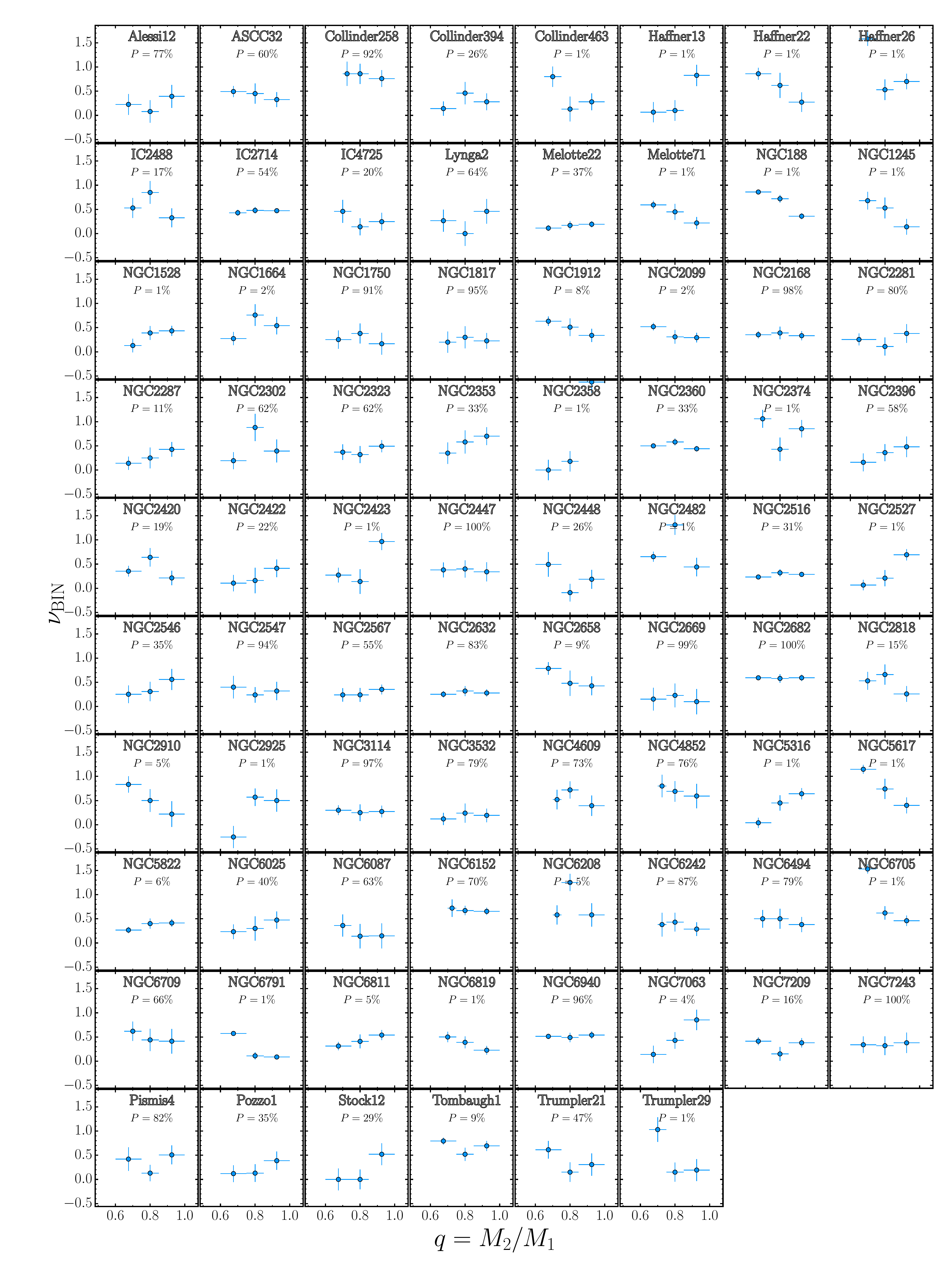}
    \caption{Mass-ratio distribution of binary stars for the 78 open clusters studied here. Clusters are sorted by name.}
    \label{fig:massratio_dist_all}
\end{figure*}

\begin{figure*}
    \centering
    \includegraphics[width=0.9\textwidth]{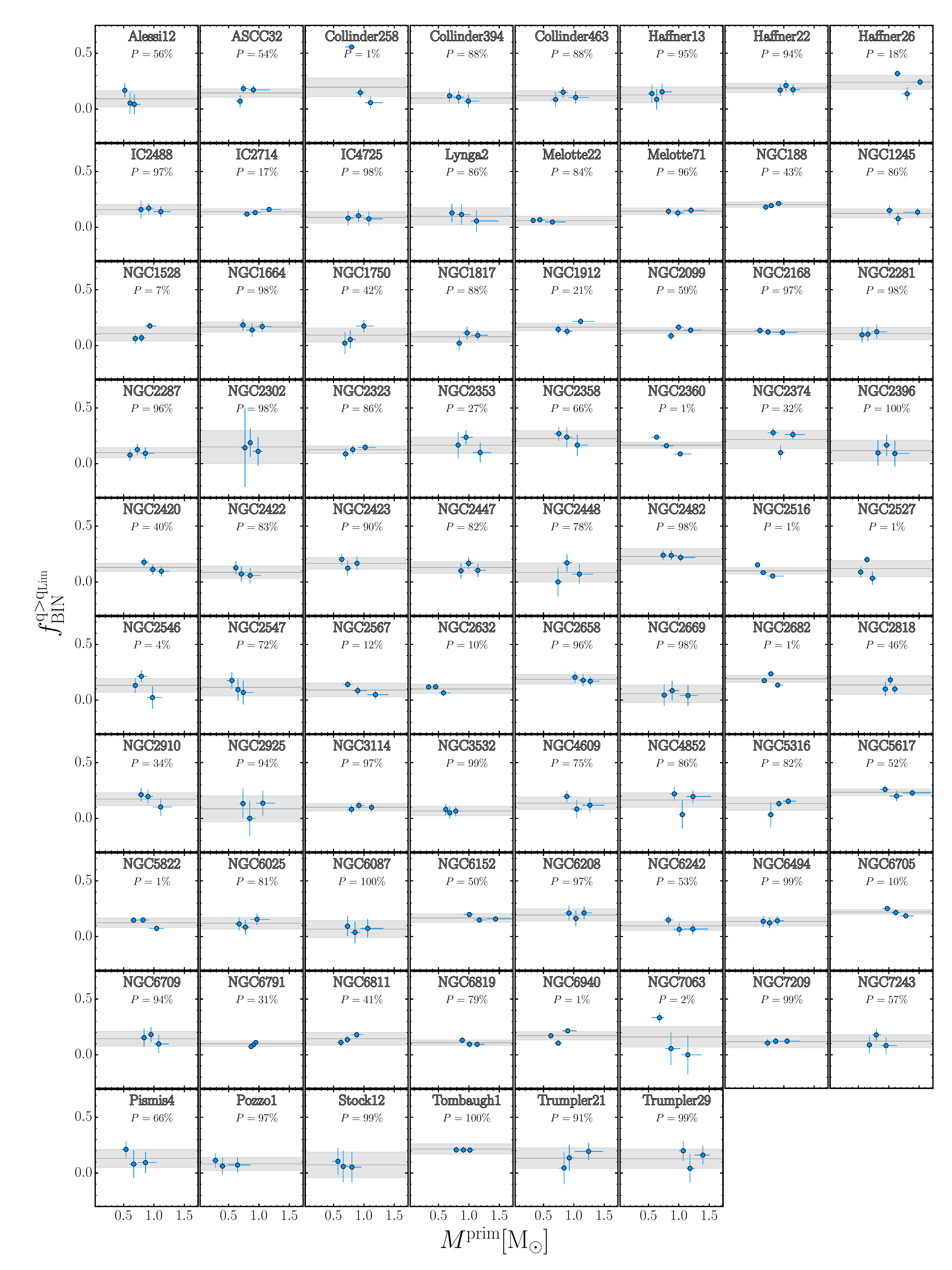}
    \caption{Mass distribution of binary stars for the 78 open clusters  studied here.}
    \label{fig:mass dist all}
\end{figure*}

\begin{figure*}
    \centering
    \includegraphics[width=0.9\textwidth]{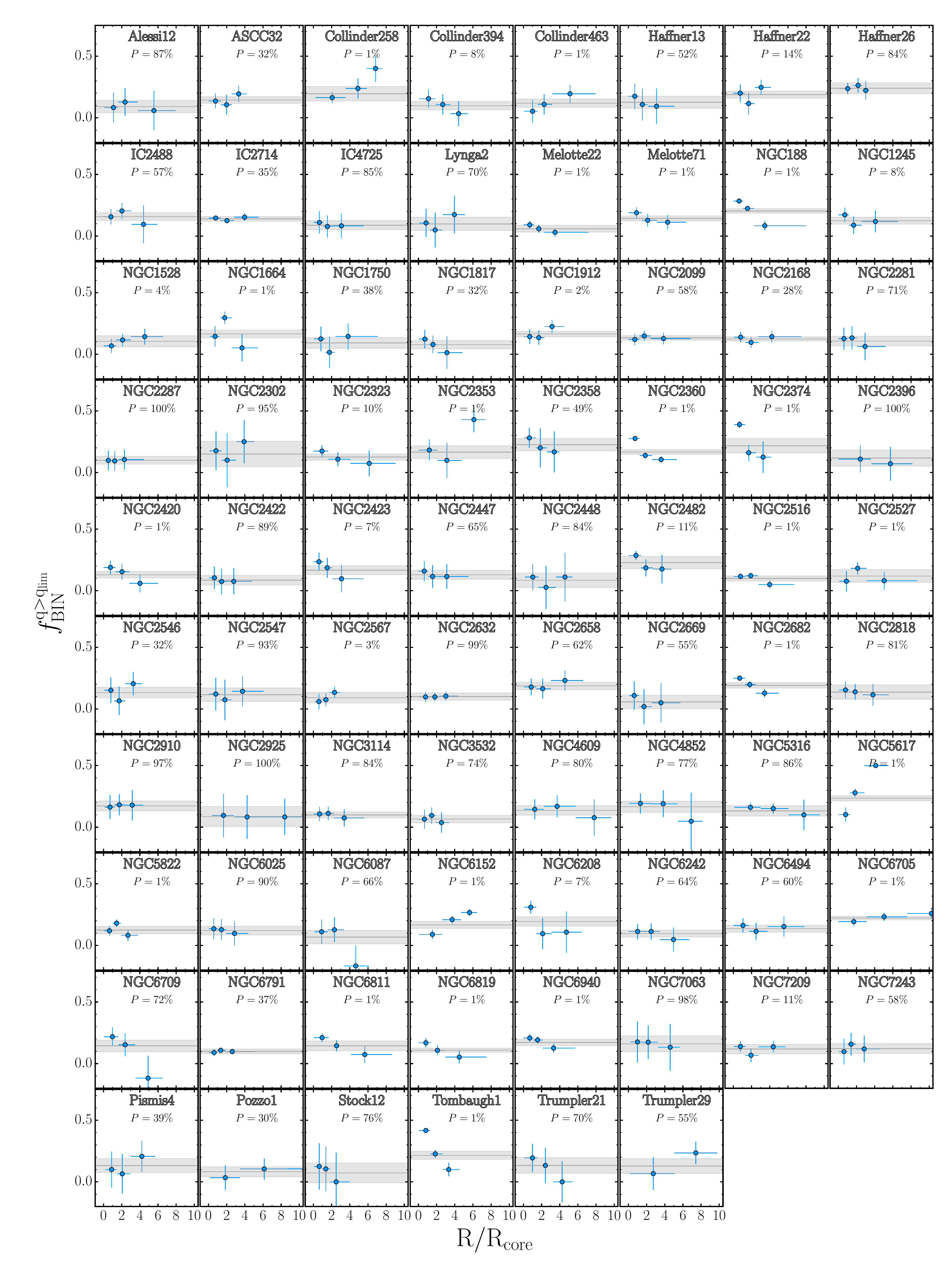}
    \caption{Radial distribution of binary stars for the 78 open clusters  studied here.}
    \label{fig:rad dist all}
\end{figure*}

Here, we show the individual binary distributions as a function of mass ratio ($q$, Fig.~\ref{fig:massratio_dist_all}), primary star mass ($M_{\rm prim}$, Fig.~\ref{fig:mass dist all}), and distance from the cluster center (\ref{fig:rad dist all}).

\section{Correlations with clusters' structural and physical properties}
\begin{figure*}
    \centering
    \includegraphics[width=0.9\textwidth]{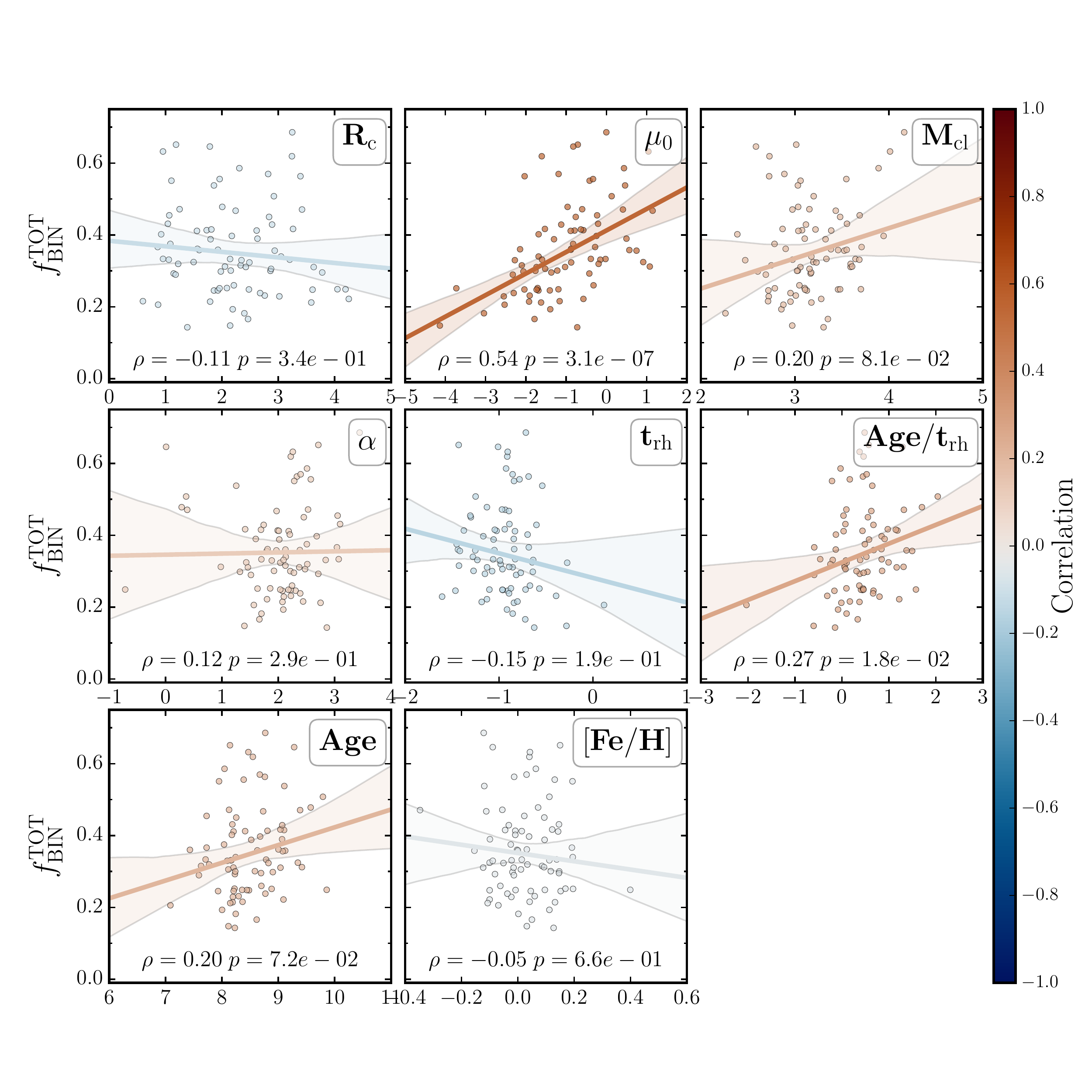}
    \caption{Correlations between total binary fraction and the physical parameters of  clusters. The marker color and line color of each panel are indicative of the correlations as shown in the right color bar. The top insets indicate the considered physical parameters, i.e., the $x$-axis, while the values of the Spearman rank coefficients and p-values are indicated in the bottom inset. Variables $\mu_0, M_{\rm cl}, t_{\rm rh}, Age/t_{\rm rh}, Age$ are in logarithm.}
    \label{fig:all corr 0}
\end{figure*}

\begin{figure*}
    \centering
    \includegraphics[width=0.9\textwidth]{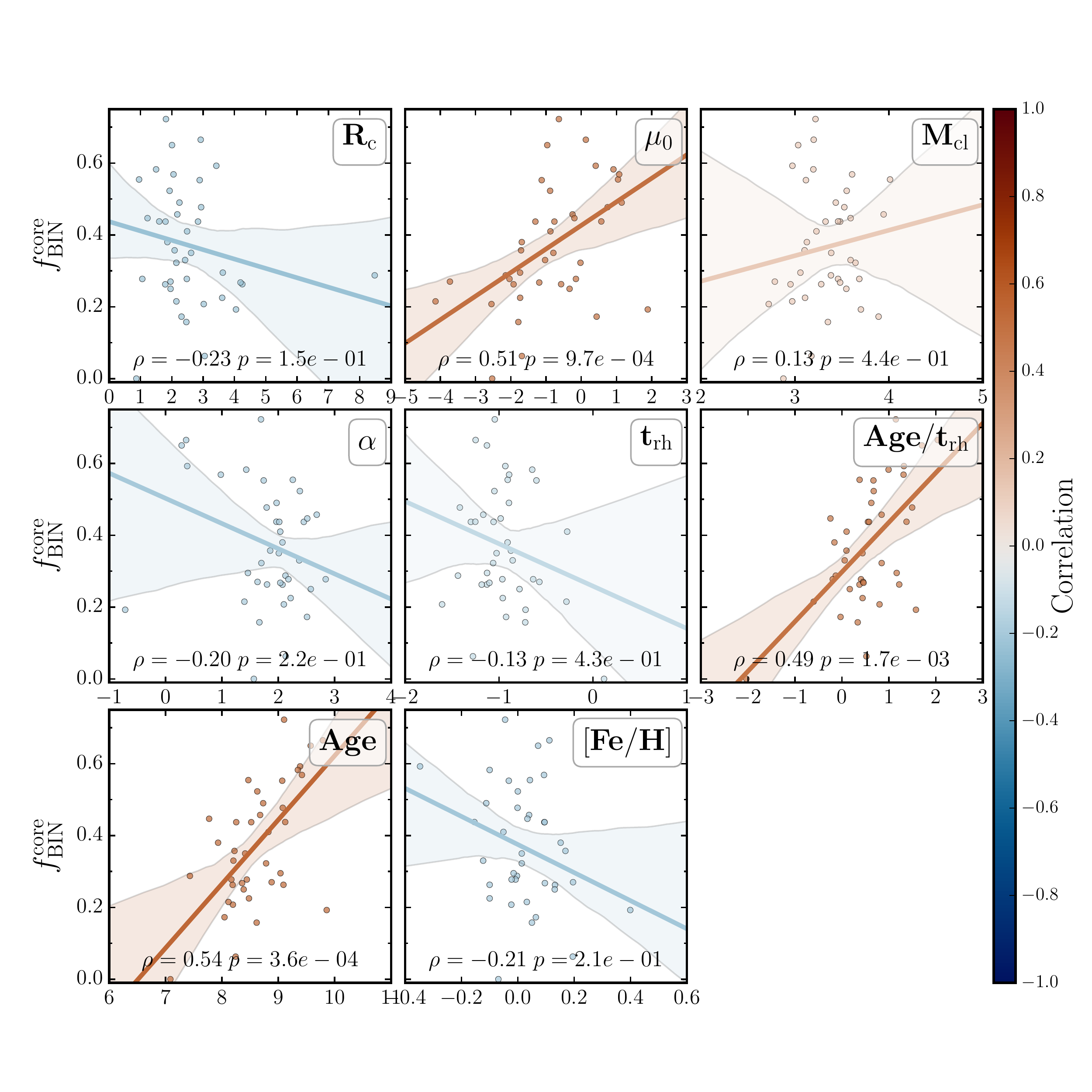}
    \caption{Correlations between binary fraction in the core and the physical parameters of  clusters. As discussed in the text, we limit the analysis of core binary fractions to clusters with more than 400 stars. The marker color and line color of each panel are indicative of the correlations as shown in the right color bar. The top insets indicate the considered physical parameters, i.e., the $x$-axis, while the values of the Spearman rank coefficients and p-values are indicated in the bottom inset. Variables $\mu_0, M_{\rm cl}, t_{\rm rh}, Age/t_{\rm rh}, Age$ are in logarithm.}
    \label{fig:all corr 1}
\end{figure*}

Figures~\ref{fig:all corr 0} and \ref{fig:all corr 1} show the correlations between total and core binary fractions, respectively. We reiterate here that core binary fractions are computed only for clusters with more than 500 stars.

\begin{table*}
    \caption{Cluster properties and density profiles. ID, right ascension, declination, proper motions, parallax, distance modulus, log(age), reddening, number of stars, differential reddening flag, core radius (arcmin) and associated error, core radius (parsecs) and associated error, central density (stars/arcmin$^2$) and (stars/parsec$^2$) and associated errors.}
    \centering
    {\tiny
    \begin{adjustbox}{angle=90}
        \begin{tabular}{lccccccccccccccccccc}
            \hline
            \hline
            Cluster &     RA &    DEC &   $\mu_{\alpha}cos\delta$ &  $\mu_\delta$ &  $\omega$ &    $(M-m)_0$ &  $log(Age)$ &   $[Fe/H]$ &   $A_V$ &   $N_{star}$ &  \texttt{DR}  & $R_{core}$ &  $\sigma R_{core}$ &  $R_{core}^{pc}$ &  $\sigma R_{core}^{pc}$ &    $\mu_0$ &  $\sigma \mu_0$ &  $\mu_0^{pc}$ &  $\sigma \mu_0^{pc}$ \\
             &     [deg] &    [deg] &  [mas/yr] &  [mas/yr] &  [mas] &  [mag]  &  [yr] &  [dex] &  &    &    & [arcmin] &  &  [pc] &  & [stars/arcmin$^2$] &   &  [stars/pc$^2$] &   \\
            &        &        &        &        &      &         &       &       &      &         &       &      &        &         &       &      &        &         \\
            \hline
            \textbf{Alessi12    } &  310.90 &     23.90 &    4.33 &   -4.65 &  1.82 &   8.65 &  8.20 &  -0.02 &  0.25 &   517 &   1 &  19.26 &  3.27 &  3.02 &  0.51 &   0.12 &  0.02 &   0.0028 &  0.0005 \\
            \textbf{ASCC32      } &  105.71 &    -26.58 &   -3.32 &    3.48 &  1.24 &   9.49 &  7.43 &  -0.00 &  0.22 &  2064 &   1 &  36.81 &  4.15 &  8.48 &  0.96 &   0.14 &  0.00 &   0.0072 &  0.0005 \\
            \textbf{Collinder258} &  186.77 &    -60.76 &   -7.07 &   -0.26 &  0.77 &  10.38 &  8.14 &   0.15 &  0.75 &   236 &   0 &   3.42 &  0.81 &  1.19 &  0.28 &   1.62 &  0.21 &   0.1949 &  0.0254 \\
            \textbf{Collinder394} &  283.06 &    -20.22 &   -1.47 &   -5.88 &  1.39 &   9.20 &  7.93 &   0.15 &  0.90 &   691 &   1 &   9.23 &  1.25 &  1.86 &  0.25 &   0.51 &  0.03 &   0.0205 &  0.0011 \\
            \textbf{Collinder463} &   26.95 &     71.77 &   -1.71 &   -0.30 &  1.14 &   9.59 &  8.24 &   0.20 &  1.10 &   679 &   1 &  12.65 &  2.34 &  3.05 &  0.56 &   0.36 &  0.07 &   0.0207 &  0.0043 \\
            \textbf{Haffner13   } &  115.24 &    -30.07 &   -6.19 &    5.89 &  1.74 &   8.74 &  7.63 &   0.09 &  0.17 &   492 &   0 &  14.30 &  1.79 &  2.33 &  0.29 &   0.30 &  0.03 &   0.0079 &  0.0007 \\
            \textbf{Haffner22   } &  123.10 &    -27.90 &   -1.64 &    2.88 &  0.33 &  12.15 &  9.39 &  -0.35 &  0.70 &   721 &   1 &   4.35 &  0.66 &  3.42 &  0.52 &   4.17 &  0.35 &   2.5763 &  0.2164 \\
            \textbf{Haffner26   } &  123.91 &    -30.85 &   -1.58 &    2.37 &  0.30 &  12.00 &  8.77 &  -0.12 &  0.76 &   703 &   1 &   4.43 &  0.82 &  3.25 &  0.60 &   1.86 &  0.27 &   0.9971 &  0.1443 \\
            \textbf{IC2488      } &  141.85 &    -57.02 &   -7.76 &    5.69 &  0.72 &  10.52 &  8.25 &   0.09 &  0.89 &   822 &   1 &   7.67 &  1.13 &  2.84 &  0.42 &   1.26 &  0.06 &   0.1734 &  0.0078 \\
            \textbf{IC2714      } &  169.38 &    -62.71 &   -7.59 &    2.66 &  0.72 &  10.44 &  8.68 &   0.04 &  1.22 &  1761 &   1 &   6.09 &  0.57 &  2.18 &  0.20 &   4.43 &  0.28 &   0.5666 &  0.0310 \\
            \hline
            \hline
        \end{tabular}
    \end{adjustbox}}
    \label{tab:tab1}
\end{table*}

\begin{table*}
    \caption{Cluster structural parameters and binary fractions. IF, MF slope, low-mass MF slope, high-mass MF slope with relative uncertainties, cluster mass, half-mass radius, hal-fmass relaxation time minimum binaries mass-ratio $q$, binary fractions and errors, blue-straggler fraction, and number. }
    \centering
    \begin{adjustbox}{rotate=90}
        \begin{tabular}{lccccccccccccccccccc}
            \hline
            \hline
            Cluster &    $\alpha_{all}$ &   $\sigma \alpha_{all}$ &    $\alpha_{low}$ &   $\sigma \alpha_{low}$ &   $\alpha_{high}$ &  $\sigma\alpha_{high}$ &   $M_{cl}$ &   $R_{\rm halfmass}$ &   $t_{\rm rh}$ &  $q_{lim}$ &  $f_{BIN}^{qlim}$ &  $\sigma f_{BIN}^{qlim}$ &  $f_{bin}^{tot}$ &  $\sigma f_{bin}^{tot}$ &  $f_{bin}^{core}$ &  $\sigma f_{bin}^{core}$ &   $f_{BSs}$ &  $\sigma f_{BSs}$ &  $N_{BSs}$ \\
             &           &           &           &           &           &           &               &   [$M_\odot$]    &     [arcmin]       &      [Myr]       &           &            &                &                 &           &           &       \\
             &           &           &           &           &           &           &               &       &           &             &           &            &                &                 &           &           &       \\
            \hline
            \hline
            \textbf{Alessi12    } &    2.10 &     0.05 &    1.41 &     0.03 &     3.76 &       0.05 &       5268 &       42.08 &   24 &  0.60 &       0.09 &        0.07 &      0.23 &       0.18 &           0.21 &            0.42 &     0.00 &      0.00 &  0 \\
            \textbf{ASCC32      } &    2.13 &     0.03 &    1.08 &     0.02 &     3.69 &       0.03 &      2417 &       73.05 &   36 &  0.60 &       0.14 &        0.04 &      0.36 &       0.10 &           0.29 &            0.20 &     0.00 &      0.00 &  0 \\
            \textbf{Collinder258} &    2.73 &     0.12 &    2.59 &     0.08 &     3.05 &      0.27 &      1032 &        6.32 &   37 &  0.70 &       0.20 &        0.08 &      0.65 &       0.28 &           1.00 &            0.42 &     0.00 &      0.00 &  0 \\
            \textbf{Collinder394} &    2.07 &     0.04 &    2.11 &     0.05 &     3.02 &      0.24 &      1340 &       22.42 &  123 &  0.60 &       0.10 &        0.05 &      0.25 &       0.12 &           0.38 &            0.30 &     0.00 &      0.00 &  0 \\
            \textbf{Collinder463} &    2.13 &     0.05 &    1.61 &     0.06 &     2.34 &      0.06 &      1496 &       26.71 &   52 &  0.65 &       0.12 &        0.05 &      0.34 &       0.14 &           0.06 &            0.40 &     0.00 &      0.00 &  0 \\
            \textbf{Haffner13   } &    2.13 &     0.06 &    0.98 &     0.03 &     0.00 &      0.00 &       560 &       20.70 &   37 &  0.60 &       0.13 &        0.07 &      0.32 &       0.18 &           0.33 &            0.34 &     0.00 &      0.00 &  0 \\
            \textbf{Haffner22   } &    0.67 &     0.01 &    0.48 &     0.09 &    -0.19 &      0.99 &       940 &        6.82 &  116 &  0.60 &       0.19 &        0.04 &      0.47 &       0.11 &           0.59 &            0.18 &     0.07 &      0.01 &  8 \\
            \textbf{Haffner26   } &    3.44 &     0.11 &    4.78 &     0.08 &     3.37 &      0.36 &     14563 &        8.96 &  193 &  0.65 &       0.24 &        0.06 &      0.69 &       0.18 &           0.43 &            0.45 &     0.00 &       0.00 &  0 \\
            \textbf{IC2488      } &    2.45 &     0.05 &    1.55 &     0.06 &     2.71 &      0.12 &      3028 &       13.26 &   50 &  0.65 &       0.16 &        0.05 &      0.45 &       0.13 &           0.44 &            0.23 &     0.00 &      0.00 &  0 \\
            \textbf{IC2714      } &    2.68 &     0.04 &    2.66 &     0.04 &     2.76 &      0.18 &      8795 &       10.48 &   67 &  0.65 &       0.14 &        0.03 &      0.40 &       0.08 &           0.46 &            0.14 &     0.00 &      0.00 &  0 \\
            \hline
        \end{tabular}
    \end{adjustbox}
    \label{tab:tab2}
\end{table*}

\bibliographystyle{aa}
\bibliography{ms}
\end{document}